\newif\ifisarxive
\isarxivetrue  % use \isarxivetrue for ARXIV, 
\ifisarxive
    \documentclass{article}
    \pdfoutput=1
    \usepackage{arxiv}

\else
    \documentclass[12pt]{article}
    \usepackage[a4paper, margin=1in]{geometry}
\fi
\usepackage[utf8]{inputenc}
\usepackage{graphicx}
\usepackage{tikz}
\usepackage{booktabs}
\usepackage{mathtools}
\usepackage{amsmath, amsthm, amssymb, amsfonts}
\usepackage{subcaption}
\usepackage{enumitem}
\usepackage{multirow}
\usepackage{float}
\usepackage{hyperref}
\usepackage[toc,page]{appendix}

\usepackage{glossaries}
\newacronym{eg}{EG}{equivalent groups}
\newacronym{nec}{NEC}{nonequivalent groups with covariates}
\newacronym{neat}{NEAT}{nonequivalent groups with anchor test}
\newacronym{see}{SEE}{standard error of equating}
\newacronym{nl}{NL}{Native Language}
\newacronym{el}{EL}{English Language}
\newacronym{gke}{GKE}{generalized kernel equating}
\newacronym{rmse}{RMSE}{root mean squared error}
\newacronym{dtm}{DTM}{difference that matters}
\newacronym{ediff}{EDIFF}{equating differences}
\newacronym{tvd}{TVD}{total variation distance}

\usepackage{setspace}
\usepackage[style = apa, sortcites = true, sorting = nyt,backend = biber]{biblatex}

\nocite{*}

\title{Sequential generalized kernel equating:\\ Providing comparable scores \\across multiple test forms with nonequivalent groups \\and differently measured covariates}

\ifisarxive
    \renewcommand{\shorttitle}{Sequential generalized kernel equating}
\fi

\author{Michaela Vařejková$^{1,2}$, Patrícia Martinková$^{1,3}$, Eva Potužníková$^{3}$\\
\small $^{1}$ Institute of Computer Science of the Czech Academy of Sciences, Prague, Czech Republic\\
\small $^{2}$ Department of Probability and Mathematical Statistics, Faculty of Mathematics and Physics, Charles University, Prague, Czech Republic\\
\small $^{3}$ Institute for Research and Development in Education, Faculty of Education, Charles University, Prague, Czech Republic\\
}

\date{}

\begin{document}

\maketitle

\begin{abstract}
Test equating using covariates may be applied to provide comparable scores from multiple test forms when no anchor items are available. However, its performance may be compromised if some of the covariates themselves are measured using different test forms. In this work, we propose sequential generalized kernel equating to account for possible differences in the distribution of covariates used in the \gls{nec} design. We evaluate the proposed approach through a simulation study within the generalized kernel equating framework.  
Results indicate that equating the covariate reduces bias in equated test scores, particularly when the covariate distributions differ, and the correlation between the covariate and the test score is strong. A real data example from a national high school leaving examination further demonstrates the practical application, and the discussion considers its potential use in linking and harmonizing scores from related psychological and behavioral measures.
\end{abstract}

% keywords

%------------------------------------------------------------
\section{Introduction}
%------------------------------------------------------------

Large-scale educational assessments often involve multiple test forms that are administered at different times or different places but are intended to measure the same underlying knowledge and skills. Multiple test forms are typically needed to ensure that all test takers are assessed under the same conditions, without those tested later having an unfair advantage of knowing the content of past administrations. Other examples of alternate versions of the same test in large-scale assessment settings include different language variants, special educational needs accommodations, matrix-sampled tests, or computer adaptive tests \parencite{depascale2020comparability}. Different test forms may also be needed to update the test items to reflect changes in the curriculum or the broader social context in which the assessment takes place \parencite{sansivieri2017review}. Since the most common use of educational assessments is to compare the performance of testees, regardless of which specific form they took, equating scores from different test forms is necessary to make the comparison fair (American Educational Research Association et al., \citeyear{aera2014standards}). Fair and accurate equating is therefore an important psychometric topic, and continued research on both the statistical tools and the computational aspects of equating remains essential \parencite{martinkova2023computational}.

A variety of test equating methods can be used, depending on the testing situation. While a common goal is to disentangle test form differences from the examinees' differences in ability, the appropriate methods are chosen according to the data collection design \parencite{vonDavier2013observed}. Two common approaches implemented in large-scale assessment programs are the \gls{eg} design which assumes that the groups of examinees taking different test forms do not differ in ability, and the \gls{neat} design, which requires the use of an anchor test, where a common set of items is administered to all test takers in addition to their specific test form to account for possible differences in ability between the groups \parencite{gonzalez2017applying, kolen2004test, sansivieri2017review, vonDavier2010statistical, vonDavier2013observed}. However, in situations where the groups cannot be considered equivalent, and the anchor test is not applicable or feasible, alternative methods must be used. One approach is to substitute the anchor test scores with covariates, for example, school grades or scores from other tests, and apply equating in \glsfirst{nec} design \parencite{wiberg2015kernel,wallin2019kernel}. The critical assumptions are that the covariates can explain differences in ability and that the conditional distribution of test scores, given the covariates, is the same across all groups. These assumptions ensure that the covariates adequately explain the differences between groups. This equating method, however, is threatened if the relationship between test scores and covariates, or the meaning of the covariates themselves, differs between the groups. This may occur, for example, when some of the covariates are measured using multiple test forms. In such cases, the underlying assumption of the same conditional distributions may be violated, which can lead to biased equating results, as discussed in \textcite{SanMartin2022NEAT}.

In this work, we provide evidence via a simulation study of how violations of the assumption of population-invariant conditional distributions may bias the results of equating. We use the \gls{gke} framework \parencite{wiberg2025generalized} due to its flexibility, minimal distributional assumptions, and suitability for incorporating covariate information. We focus on the case where one of the covariates used in the equating process is measured using multiple test forms, exploring variants with similar and different test form difficulties, as well as those with stronger and weaker correlations between the covariate and the primary test score. For the \gls{nec} design when both the construct of interest and the covariates are measured using multiple test forms, we propose applying \textit{sequential equating} to improve the accuracy of the equating process. 

The paper proceeds as follows: In Section~2, we summarize \gls{gke} with the \gls{nec} design and its underlying assumptions. We then propose and formalize the sequential generalized kernel equating approach. In Section 3, we present a simulation study to investigate the impact of assumption violation and to evaluate the proposed approach. Section 4 includes an empirical illustration. Finally, we provide a discussion and concluding remarks in Section 5.

%------------------------------------------------------------
\section{Methods}\label{sec2:methods}
%------------------------------------------------------------

\subsection{Generalized kernel equating with the \gls{nec} design}\label{sec2.1:kernel covariate equating}

\gls{gke} provides a general framework for equating under a~variety of data collection designs, including the \gls{nec} design, in which scores from different test forms are equated while accounting for covariates. The \gls{gke} framework is flexible, encompassing several traditional equating methods—such as mean, linear, and equipercentile equating—as special cases \parencite{vonDavier2004kernel}. It can be described as a sequence of five steps: presmoothing, estimation of score probabilities, continuization, equating, and evaluating the equating transformation \parencite{wiberg2025generalized}. We first introduce the notation and underlying framework that will be used throughout the article. 

We denote the random variables representing the total scores from two different test forms as $X$ and $Y$, where both forms are designed to measure the same construct, and the vector of covariates as $\boldsymbol{C}$. The covariate vector $\boldsymbol{C}$ can take values $\boldsymbol{c_l}$, $l=1, \dots ,L$, where $L$ represents the number of possible covariate categories or combinations of covariate values. If continuous covariates are included in $\boldsymbol{C}$, they are discretized into a finite number of intervals so that the number of covariate categories $L$ remains finite.
Furthermore, we consider two populations of students, populations $P$ and $Q$. Generally, we assume that these populations differ in their ability. Our target population of interest $T$ is assumed to be a mixture of these two populations:
\begin{align}\label{target_population}
    T = \omega P + (1 - \omega )Q, 
\end{align} 
where $0 \leq \omega \leq 1$ is a~constant representing the mixture weight between the populations, commonly determined based on the relative sample sizes of the populations. This mixture formulation is symbolic, representing how score distributions for $T$ are derived. 

We further assume that a group of students who took test form X (with total scores represented by the random variable $X$) was randomly selected from the population $P$, and a group of students who took test form Y (with total scores represented by the random variable $Y$) was randomly selected from the population $Q$. The joint probability distributions of $(X, \boldsymbol{C})$ and $(Y, \boldsymbol{C})$, in populations $P$ and $Q$, respectively, are defined as

\begin{equation}
\begin{aligned}\label{pst p a q}
p_{jl} = P(X = x_j, \boldsymbol{C} = \boldsymbol{c_l} |P), \; \;
q_{kl} = P(Y = y_k, \boldsymbol{C} = \boldsymbol{c_l} |Q),
\end{aligned}
\end{equation}

\noindent
and $\mathbf{p} = \left( p_{jl}\right)_{j=1,\ldots,J;\,l=1,\ldots,L}^\top$, $\mathbf{q} = \left(q_{kl}\right)_{k=1,\ldots,K;\,l=1,\ldots,L}^\top$ . 

\paragraph{Presmoothing. } The first step in \gls{gke} is presmoothing, which aims to reduce sample irregularities in the empirical score distributions and stabilize the estimated probabilities $p_{jl}$ and $q_{kl}$. Presmoothing is not mandatory, especially for very large sample sizes, but it is often used in practice. It can be carried out using a variety of methods, with polynomial log-linear models being commonly employed in kernel equating \parencite{holland_thayer_2000}.

\paragraph{Estimation of score probabilities. } Since $T$ is a mixture of populations $P$ and $Q$ as stated in Equation~\eqref{target_population}, we can express vectors of score probabilities $\textbf{r} = (r_1, \dots, r_J)^\top$ and $\textbf{s} = (s_1, \dots, s_K)^\top$, and the vector of covariate probabilities $\textbf{t} = (t_1, \dots, t_L)^\top$ as follows:

\begin{equation}
\begin{aligned}
\label{psti r, s a t}
r_j &= P(X = x_j | T) = \omega r_{Pj} + (1 - \omega)r_{Qj}, \\
s_k &= P(Y = y_k | T) = \omega s_{Pk} + (1 - \omega)s_{Qk}, \\
t_l &= P(\boldsymbol{C} = \boldsymbol{c_l} | T) = \omega t_{Pl} + (1 - \omega)t_{Ql},
\end{aligned}
\end{equation}
where
\begin{equation}
\begin{aligned}\label{psti v P a Q}
r_{Pj} &= P(X = x_j | P) = \sum_{l=1}^L p_{jl}, \enspace \enspace &r_{Qj} &= P(X = x_j | Q), \\
s_{Qk} &= P(Y = y_k | Q) = \sum_{l=1}^L q_{kl}, \enspace \enspace &s_{Pk} &= P(Y = y_k | P), \\
t_{Pl} &= P(\boldsymbol{C} = \boldsymbol{c_l} | P) = \sum_{j=1}^J p_{jl}, \enspace \enspace
&t_{Ql} &= P(\boldsymbol{C} = \boldsymbol{c_l} | Q) = \sum_{k=1}^K q_{kl}.
\end{aligned}
\end{equation}

\noindent Since test form~$X$ was not administered to population $Q$ and test form $Y$ was not administered to population $P$, probabilities $r_{Qj}$ and $s_{Pk}$ cannot be estimated from the data. Therefore, to express probabilities $r_j$ and $s_k$, additional assumptions need to be added. The following assumption is typically used in test equating methods for \gls{nec} design \parencite{wiberg2015kernel} to identify unobserved score distributions and express the score probabilities in the target population $T$. \\

\noindent
\textbf{Assumption of population-invariant conditional distributions}:\\
For all populations $T$ (regardless of the value of $\omega$), the conditional distribution of $X$ given $\boldsymbol{C}$ is invariant with respect to the population. This can be expressed as
\begin{align}
r_Q(x_j | \boldsymbol{c_l}) = r_P (x_j | \boldsymbol{c_l}),\label{eq:assumption1}
\end{align}
where $r_Q(x_j | \boldsymbol{c_l})$ and $r_P(x_j | \boldsymbol{c_l})$ are conditional probabilities defined as  
$r_Q (x_j |\boldsymbol{c_l}) = P(X = x_j | \boldsymbol{C} = \boldsymbol{c_l}, Q), \enspace r_P (x_j |\boldsymbol{c_l}) = P(X = x_j | \boldsymbol{C} = \boldsymbol{c_l}, P).$

\noindent
Additionally, for all populations $T$ (regardless of the value of $\omega$), the conditional distribution of $Y$ given $\boldsymbol{C}$ is invariant with respect to the population. This can be expressed as
\begin{align}
s_Q(y_k | \boldsymbol{c_l}) = s_P (y_k | \boldsymbol{c_l}),\label{eq:assumption2}
\end{align}
where $s_P(y_k | \boldsymbol{c_l})$ and $s_Q(y_k | \boldsymbol{c_l})$ are conditional probabilities defined as 
$s_Q (y_k |\boldsymbol{c_l}) = P(Y = y_k | \boldsymbol{C} = \boldsymbol{c_l}, Q), \enspace  s_P (y_k |\boldsymbol{c_l}) = P(Y = y_k | \boldsymbol{C} = \boldsymbol{c_l}, P).$ 

The assumption of population-invariant conditional distributions, specified in~\eqref{eq:assumption1} and \eqref{eq:assumption2}, means that once we account for the covariates, the distribution of the test scores is the same in both populations. For example, if students with a particular covariate profile have a certain distribution of scores on test X in population $P$, then the students with the same covariate profile in population $Q$ are assumed to follow the same distribution on test X, even though test X was not administered in $Q$. The assumption then allows us to express the probabilities $r_{Qj}$ %(probabilities of test scores $X$ in population $Q$) 
and  $s_{Pk}$ %(probabilities of test scores $Y$ in population $P$)
as 
\begin{equation}
\begin{aligned}
    r_{Qj} &= \sum_{l=1}^L r_Q (x_j | \boldsymbol{c_l}) t_{Ql} 
    =  \sum_{l=1}^L r_P (x_j | \boldsymbol{c_l}) t_{Ql}
    = \sum_{l=1}^L \frac{p_{jl}}{t_{Pl}} t_{Ql}, \\
    s_{Pk} &= \sum_{l=1}^L s_P (y_k | \boldsymbol{c_l}) t_{Pl} 
    =  \sum_{l=1}^L s_Q (y_k | \boldsymbol{c_l}) t_{Pl}
    = \sum_{l=1}^L \frac{q_{kl}}{t_{Ql}} t_{Pl}
\end{aligned}
\end{equation}
and inserting it into (\ref{psti r, s a t}), we get

\begin{equation}
    \begin{aligned}
        r_j = \sum_{l=1}^L \left[ \omega + \frac{(1 - \omega) t_{Ql}}{t_{Pl}} \right] p_{jl}, \; \;
        s_k = \sum_{l=1}^L \left[ (1 - \omega) + \frac{\omega t_{Pl}}{t_{Ql}} \right] q_{kl}.
    \end{aligned}
\end{equation}

\paragraph{Continuization. }In the next step of \gls{gke}, the cumulative distribution functions of test scores are typically continuized, since test scores are usually discrete and continuization ensures that their inverse is well-defined for constructing the equating function. Kernel functions $K_V$ and $K_W$ are used to spread the probability mass of the discrete scores, producing continuous distribution functions $F_{\widetilde{X}}$ and $G_{\widetilde{Y}}$ (\cite{vonDavier2004kernel}):
\begin{equation}
    \begin{aligned}
        F_{\widetilde{X}} (x) = \sum_{j=1}^J r_j K_V \left( \frac{x - a_X x_j - \left( 1 - a_X \right) \mu_X}{a_X h_X} \right) ,
    \end{aligned}
\end{equation}
and
\begin{equation}
    \begin{aligned}
        G_{\widetilde{Y}} (y) = \sum_{k=1}^K s_k K_W \left( \frac{y - a_Y y_k - \left( 1 - a_Y \right) \mu_Y}{a_Y h_Y} \right) .
    \end{aligned}
\end{equation}

\paragraph{Equating. }With continuized distribution functions $F_{\widetilde{X}}$ and $G_{\widetilde{Y}}$ the equating function can be defined as 
     $\varphi_{GKE} (x) = G_{\widetilde{Y}}^{-1} \left( F_{\widetilde{X}} (x) \right).$

\paragraph{Evaluating the equating transformation}. The final step in the \gls{gke} framework is to evaluate the estimated equating transformation. 
A commonly used measure of precision is the \gls{see}, defined as 
\begin{equation}
SEE (x) = \sqrt{\mbox{var}  ( \widehat{ \varphi}_{GKE} (x) )}, 
\end{equation}
where $\widehat{ \varphi}_{GKE} (x)$ denotes the estimated equating function at score point $x$. 

In simulation studies, where the true equating transformation is known, the accuracy of the estimated equating function can also be evaluated in terms of bias and \gls{rmse}. Let $\hat{e}^{(r)}(s)$ denote the equated score at score point $s$ in replication $r$, let $e(s)$ denote the true equating transformation, and let $R$ denote the number of replications. Bias is defined as
\begin{equation}
Bias(s) = \frac{1}{R} \sum_{r=1}^{R} \left( \hat{e}^{(r)}(s) - e(s) \right).
\end{equation}

The \gls{rmse} combines systematic and random error and is defined as
\begin{equation}
RMSE(s) = \sqrt{Bias(s)^2 + SEE(s)^2}.
\end{equation}

When two equating transformations are compared, their discrepancy can be summarized using the equating difference \parencite{Liang2014CrossValidationKernelEquating},
\begin{equation}\label{ediff}
EDIFF(s) = \frac{1}{R} \sum_{r=1}^{R} 
\left| \hat{e}^{(r)}_1(s) - \hat{e}^{(r)}_2(s) \right|,
\end{equation}
where $\hat{e}^{(r)}_1(s)$ and $\hat{e}^{(r)}_2(s)$ denote the equated scores produced by two equating approaches in replication $r$. 
The practical relevance of these differences can be evaluated using the \gls{dtm}, which provides a criterion for judging whether differences between equating functions are large enough to be meaningful in score-point units. Following \textcite{DoransFeigenbaum1994}, \gls{dtm} was defined as one half of a reported score point.

\subsection{Sequential generalized kernel equating}\label{sec:RKCE}

In practical applications, the assumption of population-invariant conditional distributions stated in Equations~\eqref{eq:assumption1} and \eqref{eq:assumption2} may not always hold, potentially compromising the accuracy of the equating process. A common scenario in which this violation occurs is when covariates used to adjust for group differences in the \gls{nec} design are themselves measured on non-comparable scales across populations, meaning that the same value may correspond to different levels of the underlying trait in each group. This situation arises whenever the measurement process differs between populations. For example, this may occur for continuous covariates such as test scores when different test forms have different difficulties, but it may also arise for categorical covariates such as grades if grading standards differ across groups. In this paper, we illustrate the approach using a continuous test-score covariate. In such cases, the covariate may itself be affected by test form differences and therefore does not represent a purely baseline, pre-treatment variable. From a causal inference perspective, conditioning on post-treatment variables can introduce bias, as these variables may capture not only baseline differences in ability but also effects of the test form itself \parencite{RosenbaumRubin1983, Pearl2009}. To address this challenge, we propose sequential generalized kernel equating — an extension within the \gls{gke} framework — as a two-step approach that first equates the covariate distribution before applying \gls{gke} to the primary test scores. The proposed approach extends \gls{gke} by introducing a preliminary equating step for covariates measured on non-comparable scales, without altering the underlying \gls{gke} estimators or the \gls{nec} design structure. The transformed covariate is then used in the subsequent equating of the primary test scores, which is performed within the \gls{gke} framework under the \gls{nec} design. This additional equating step aims to make the covariate scores comparable across groups, increasing the chance of satisfying the assumption of population-invariant conditional distributions and, consequently, improving the accuracy of the main equating process. It can be viewed as addressing this issue by reducing form effects in the covariate prior to its use in equating. This perspective is consistent with the potential outcomes framework, where identification typically relies on conditioning on pre-treatment covariates to ensure comparability between groups \parencite{RosenbaumRubin1983, ImbensRubin2015}.\\

\noindent
\textbf{Step 1: Equating of the covariate}\\
To account for scale differences in covariates, we introduce an intermediate equating step where the covariate $C$ is first equated between populations $P$ and $Q$. Let $C^*$ denote the equated covariate such that
\begin{equation}
    \begin{aligned}\label{transformed_covariate}
    C^* = \varphi_{E} (C), 
    \end{aligned}
\end{equation}
where $\varphi_{E}$ represents the equating function derived using a suitable equating method selected according to the characteristics of the covariate:
\begin{equation}
\begin{aligned}
     \varphi_{E} (c) = G_{\widetilde{C}}^{-1} \left( F_{\widetilde{C}} (c) \right).
    \end{aligned}
\end{equation}
Here, $F_{\widetilde{C}}$ and $G_{\widetilde{C}}$ are the continuous cumulative distribution functions of covariate $C$ in populations $P$ and $Q$, respectively. The equating function maps covariate values in population $Q$ onto the corresponding distribution in population $P$, ensuring comparability.\\
In practice, the transformation in Equation~\eqref{transformed_covariate} is applied to the observed covariate values in population $Q$. When the covariate is continuous, the transformed values $C^*$ are subsequently discretized into the same covariate categories $c_l$ used for the covariate in population P. Based on these categorized values, a joint frequency table of the test scores and the transformed covariate is constructed. The transformation of the covariate in population $Q$ leads to joint probabilities
\begin{equation}
\begin{aligned}\label{pst p a q equated}
q_{kl}^* = P(Y = y_k, C^* = c_l |Q).
\end{aligned}
\end{equation}
These probabilities replace $q_{kl}$ in the original \eqref{pst p a q} formulation.\\

\noindent
\textbf{Step 2: Primary score equating}\\
With the transformed covariate $C^*$, we proceed with the primary score equating using the \gls{gke} framework under the \gls{nec} design, as described in Section~\ref{sec2.1:kernel covariate equating} (hereafter referred to as the standard \gls{gke}), to equate the test scores $X$ and $Y$. We consistently work with quantities marked by an asterisk to indicate their dependence on the equated covariate. Specifically, the score probabilities are calculated as

\begin{equation}
    \begin{aligned}
        r_j^* &= \sum_{l=1}^L \left[ \omega + \frac{(1 - \omega) t_{Ql}^*}{t_{Pl}} \right] p_{jl}, \\
        s_k^* &= \sum_{l=1}^L \left[ (1 - \omega) + \frac{\omega t_{Pl}}{t_{Ql}^*} \right] q_{kl}^*,
    \end{aligned}
\end{equation}
where $t_{Pl}$ and $t_{Ql}^*$ are the marginal probabilities of the covariate categories in populations $P$ and $Q$, respectively.\\

\noindent
The probabilities $r_j^*$ and $s_k^*$ are then used to form the smoothed cumulative distribution functions of test scores

\begin{equation}
    \begin{aligned}
        F_{\widetilde{X}}^* (x) = \sum_{j=1}^J r_j^* K_V \left( \frac{x - a_X x_j - \left( 1 - a_X \right) \mu_X}{a_X h_X} \right) , \\
        G_{\widetilde{Y}}^* (y) = \sum_{k=1}^K s_k^* K_W \left( \frac{y - a_Y y_k - \left( 1 - a_Y \right) \mu_Y}{a_Y h_Y} \right) ,
    \end{aligned}
\end{equation}

leading to the final equating function

\begin{equation}\label{eq:RKCE}
    \begin{aligned}
     \varphi_{seq.\ GKE} (x) = {G_{\widetilde{Y}}^{*}}^{-1} \left( F_{\widetilde{X}}^* (x) \right).
    \end{aligned}
\end{equation}

%------------------------------------------------------------
\section{Simulation study}

To evaluate the proposed sequential \gls{gke} approach, we conducted a simulation study using the \gls{nec} design. We consider two nonequivalent groups with observed scores from test forms X and Y, and three covariates, one of them representing a score from another test, which was itself measured by two different forms assigned to the two groups. For the simulation, we primarily consider cases where the test forms X and Y have the same difficulty, while the test forms associated with the covariate (i.e., the other test score) may differ in difficulty. To better understand the performance of the proposed approach under a wider range of conditions, we also include cases with different difficulty levels for test forms X and Y. We evaluate bias, standard errors and \gls{rmse} in equated scores introduced by not equating the covariate representing the other test score under multiple scenarios varying in a) difference in test form difficulty, b) correlation between the test score and the covariate representing the other test score, and c) sample size.    

\subsection{Simulation design}

The design of the simulation study was inspired by real-world data available in the context of a national upper secondary school leaving examination, taking into account sample sizes, covariate distributions, and other relevant specifications. An example using real data from the exam is presented in the next section. The covariates consisted of two binary variables representing attended school type (academic or non-academic) and exam attempt (first or repeated), and one continuous variable representing a score from another test. As in the real-world situation, we assume that academic school students and first-time test takers achieve higher scores with the same test form difficulties and that the two test scores are positively correlated, while acknowledging that the available covariates can explain only a part of the total test score variance.

\begin{figure}[h!]
     \tikzset{every picture/.style={line width=0.75pt}} %set default line width to 0.75pt        
\scalebox{0.45}{
\begin{tikzpicture}[x=0.75pt,y=0.75pt,yscale=-3.0,xscale=3.0]
%uncomment if require: \path (0,5632); %set diagram left start at 0, and has height of 5632

%Shape: Ellipse [id:dp8871442612185106] 
\draw   (160,5150) .. controls (160,5138.95) and (168.95,5130) .. (180,5130) .. controls (191.05,5130) and (200,5138.95) .. (200,5150) .. controls (200,5161.05) and (191.05,5170) .. (180,5170) .. controls (168.95,5170) and (160,5161.05) .. (160,5150) -- cycle ;
%Shape: Ellipse [id:dp46122147749852493] 
\draw   (260,5150) .. controls (260,5138.95) and (268.95,5130) .. (280,5130) .. controls (291.05,5130) and (300,5138.95) .. (300,5150) .. controls (300,5161.05) and (291.05,5170) .. (280,5170) .. controls (268.95,5170) and (260,5161.05) .. (260,5150) -- cycle ;
%Straight Lines [id:da020477652009631253] 
\draw    (212,5150) -- (248,5150) ;
\draw [shift={(250,5150)}, rotate = 180] [color={rgb, 255:red, 0; green, 0; blue, 0 }  ][line width=0.75]    (10.93,-3.29) .. controls (6.95,-1.4) and (3.31,-0.3) .. (0,0) .. controls (3.31,0.3) and (6.95,1.4) .. (10.93,3.29)   ;
\draw [shift={(210,5150)}, rotate = 0] [color={rgb, 255:red, 0; green, 0; blue, 0 }  ][line width=0.75]    (10.93,-3.29) .. controls (6.95,-1.4) and (3.31,-0.3) .. (0,0) .. controls (3.31,0.3) and (6.95,1.4) .. (10.93,3.29)   ;
%Shape: Rectangle [id:dp3697837658001646] 
\draw   (90,5102) -- (120,5102) -- (120,5122) -- (90,5122) -- cycle ;
%Shape: Rectangle [id:dp8455726603502608] 
\draw   (340,5102) -- (370,5102) -- (370,5122) -- (340,5122) -- cycle ;
%Shape: Ellipse [id:dp8039031844644746] 
\draw  [color={rgb, 255:red, 0; green, 60; blue, 150 }  ,draw opacity=1 ] (60,5185) .. controls (60,5176.72) and (66.72,5170) .. (75,5170) .. controls (83.28,5170) and (90,5176.72) .. (90,5185) .. controls (90,5193.28) and (83.28,5200) .. (75,5200) .. controls (66.72,5200) and (60,5193.28) .. (60,5185) -- cycle ;
%Shape: Ellipse [id:dp5433580599418469] 
\draw  [color={rgb, 255:red, 0; green, 60; blue, 150 }  ,draw opacity=1 ] (370,5185) .. controls (370,5176.72) and (376.72,5170) .. (385,5170) .. controls (393.28,5170) and (400,5176.72) .. (400,5185) .. controls (400,5193.28) and (393.28,5200) .. (385,5200) .. controls (376.72,5200) and (370,5193.28) .. (370,5185) -- cycle ;
%Shape: Rectangle [id:dp9009099571547728] 
\draw   (110,5140) -- (140,5140) -- (140,5160) -- (110,5160) -- cycle ;
%Shape: Rectangle [id:dp1614412958462217] 
\draw   (320,5140) -- (350,5140) -- (350,5160) -- (320,5160) -- cycle ;
%Curve Lines [id:da8989137995196382] 
\draw    (120,5110) .. controls (143.04,5106.14) and (158.78,5114.02) .. (169.06,5131.36) ;
\draw [shift={(170,5133)}, rotate = 240.99] [fill={rgb, 255:red, 0; green, 0; blue, 0 }  ][line width=0.08]  [draw opacity=0] (12,-3) -- (0,0) -- (12,3) -- cycle    ;
%Curve Lines [id:da4641367582393944] 
\draw    (291.95,5131.19) .. controls (300.95,5114.8) and (317.56,5105.19) .. (340,5110) ;
\draw [shift={(291,5133)}, rotate = 296.59] [fill={rgb, 255:red, 0; green, 0; blue, 0 }  ][line width=0.08]  [draw opacity=0] (12,-3) -- (0,0) -- (12,3) -- cycle    ;
%Straight Lines [id:da4723492931803539] 
\draw    (140,5150) -- (158,5150) ;
\draw [shift={(160,5150)}, rotate = 180] [fill={rgb, 255:red, 0; green, 0; blue, 0 }  ][line width=0.08]  [draw opacity=0] (12,-3) -- (0,0) -- (12,3) -- cycle    ;
%Straight Lines [id:da4332885584116406] 
\draw    (302,5150) -- (320,5150) ;
\draw [shift={(300,5150)}, rotate = 0] [fill={rgb, 255:red, 0; green, 0; blue, 0 }  ][line width=0.08]  [draw opacity=0] (12,-3) -- (0,0) -- (12,3) -- cycle    ;
%Straight Lines [id:da6173146749818236] 
\draw    (100,5122) -- (80.77,5168.15) ;
\draw [shift={(80,5170)}, rotate = 292.62] [fill={rgb, 255:red, 0; green, 0; blue, 0 }  ][line width=0.08]  [draw opacity=0] (12,-3) -- (0,0) -- (12,3) -- cycle    ;
%Straight Lines [id:da0969483328677685] 
\draw    (360,5122) -- (379.23,5168.15) ;
\draw [shift={(380,5170)}, rotate = 247.38] [fill={rgb, 255:red, 0; green, 0; blue, 0 }  ][line width=0.08]  [draw opacity=0] (12,-3) -- (0,0) -- (12,3) -- cycle    ;
%Straight Lines [id:da11278533312865535] 
\draw    (110,5150) -- (87.39,5173.56) ;
\draw [shift={(86,5175)}, rotate = 313.83] [fill={rgb, 255:red, 0; green, 0; blue, 0 }  ][line width=0.08]  [draw opacity=0] (12,-3) -- (0,0) -- (12,3) -- cycle    ;
%Straight Lines [id:da08707908902604045] 
\draw    (350,5150) -- (372.61,5173.56) ;
\draw [shift={(374,5175)}, rotate = 226.17] [fill={rgb, 255:red, 0; green, 0; blue, 0 }  ][line width=0.08]  [draw opacity=0] (12,-3) -- (0,0) -- (12,3) -- cycle    ;
%Curve Lines [id:da11647868057878841] 
\draw [color={rgb, 255:red, 0; green, 60; blue, 150 }  ,draw opacity=1 ] [dash pattern={on 4.5pt off 4.5pt}]  (90,5190) .. controls (118.32,5198.39) and (148.82,5194.63) .. (170.04,5169.18) ;
\draw [shift={(171,5168)}, rotate = 128.71] [fill={rgb, 255:red, 0; green, 60; blue, 150 }  ,fill opacity=1 ][line width=0.08]  [draw opacity=0] (12,-3) -- (0,0) -- (12,3) -- cycle    ;
%Curve Lines [id:da72105326320853] 
\draw [color={rgb, 255:red, 0; green, 60; blue, 150 }  ,draw opacity=1 ] [dash pattern={on 4.5pt off 4.5pt}]  (290.65,5169.77) .. controls (312.27,5192.43) and (341.99,5200.25) .. (371,5190) ;
\draw [shift={(289,5168)}, rotate = 47.84] [fill={rgb, 255:red, 0; green, 60; blue, 150 }  ,fill opacity=1 ][line width=0.08]  [draw opacity=0] (12,-3) -- (0,0) -- (12,3) -- cycle    ;
%Rounded Rect [id:dp5037953245481669] 
\draw   (20,5154) .. controls (20,5151.79) and (21.79,5150) .. (24,5150) -- (36,5150) .. controls (38.21,5150) and (40,5151.79) .. (40,5154) -- (40,5168.48) .. controls (40,5170.69) and (38.21,5172.48) .. (36,5172.48) -- (24,5172.48) .. controls (21.79,5172.48) and (20,5170.69) .. (20,5168.48) -- cycle ;
%Straight Lines [id:da07449836072289273] 
\draw  [dash pattern={on 0.84pt off 2.51pt}]  (40,5160) -- (58.59,5178.59) ;
\draw [shift={(60,5180)}, rotate = 225] [color={rgb, 255:red, 0; green, 0; blue, 0 }  ][line width=0.75]    (10.93,-3.29) .. controls (6.95,-1.4) and (3.31,-0.3) .. (0,0) .. controls (3.31,0.3) and (6.95,1.4) .. (10.93,3.29)   ;
%Rounded Rect [id:dp9540041923758688] 
\draw   (420,5154) .. controls (420,5151.79) and (421.79,5150) .. (424,5150) -- (436,5150) .. controls (438.21,5150) and (440,5151.79) .. (440,5154) -- (440,5168.48) .. controls (440,5170.69) and (438.21,5172.48) .. (436,5172.48) -- (424,5172.48) .. controls (421.79,5172.48) and (420,5170.69) .. (420,5168.48) -- cycle ;
%Straight Lines [id:da8852841610609081] 
\draw  [dash pattern={on 0.84pt off 2.51pt}]  (420,5160) -- (401.41,5178.59) ;
\draw [shift={(400,5180)}, rotate = 315] [color={rgb, 255:red, 0; green, 0; blue, 0 }  ][line width=0.75]    (10.93,-4.9) .. controls (6.95,-2.3) and (3.31,-0.67) .. (0,0) .. controls (3.31,0.67) and (6.95,2.3) .. (10.93,4.9)   ;

% Text Node
\draw (174,5145) node [anchor=north west][inner sep=0.75pt]    {\Huge $X$};
% Text Node
\draw (275,5146) node [anchor=north west][inner sep=0.75pt]    {\Huge $Y$};
% Text Node
\draw (97,5106) node [anchor=north west][inner sep=0.75pt]    {\Huge $C_{2}$};
% Text Node
\draw (347,5106) node [anchor=north west][inner sep=0.75pt]    {\Huge $C_{2}$};
% Text Node
\draw (67,5179) node [anchor=north west][inner sep=0.75pt]    {\Huge $C_{3}$};
% Text Node
\draw (377,5179) node [anchor=north west][inner sep=0.75pt]    {\Huge $C_{3}$};
% Text Node
\draw (117,5144) node [anchor=north west][inner sep=0.75pt]    {\Huge $C_{1}$};
% Text Node
\draw (327,5144) node [anchor=north west][inner sep=0.75pt]    {\Huge $C_{1}$};
% Text Node
\draw (24,5156.48) node [anchor=north west][inner sep=0.75pt]  [color={rgb, 255:red, 155; green, 155; blue, 155 }  ,opacity=1 ]  {\Huge $\mathcal{N}$};
% Text Node
\draw (424,5156.48) node [anchor=north west][inner sep=0.75pt]  [color={rgb, 255:red, 155; green, 155; blue, 155 }  ,opacity=1 ]  {\Huge $\mathcal{N}$};

\end{tikzpicture}
}
                  \caption{Diagram illustrating the design of the simulation study.}
                  \label{fig:SimulationDiagram}
			\end{figure}
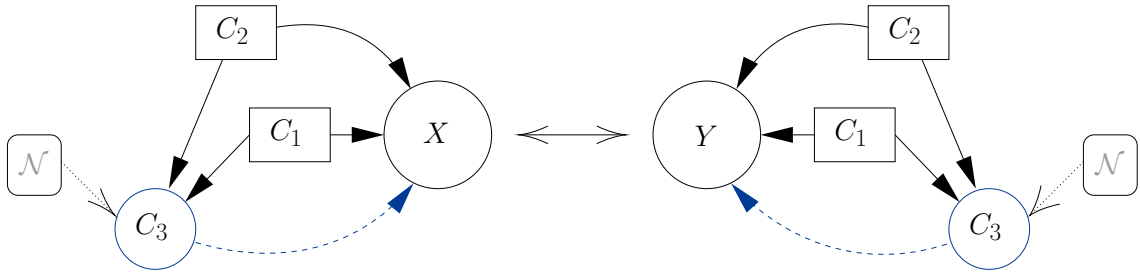

The data generation process was guided by our aim to assess the effects of different violations of the assumption of population-invariant conditional distributions on the resulting equated test scores. In some simulation scenarios, the continuous covariate corresponding to another test score was generated assuming both tests had the same difficulty. These scenarios satisfied the assumption of population-invariant conditional distributions in both groups, as defined in Equations \eqref{eq:assumption1} and \eqref{eq:assumption2}. In other scenarios, the continuous covariate was generated assuming different test difficulties, thus violating this assumption. 
The general design of the simulation study is depicted in Figure~\ref{fig:SimulationDiagram}. \\

\vspace{1em}

The data were generated as follows:

\begin{enumerate}
    \item  Initially, we generated two binary variables. The first variable, denoted as $C_1$, represents the attended school type, while the second variable, denoted as $C_2$, represents the exam attempt. These variables were simulated from a multivariate binary distribution, with population-specific marginal probabilities and odds ratio matrices \parencite{Lee1993RandomBinaryDeviates}. For population~P, the distribution was defined by marginal probabilities 
\[
p^{(P)} = (0.300,\, 0.800)
\] 
and odds ratio matrix 
\[
\begin{bmatrix}
 - & 8 \\
8 & -
\end{bmatrix},
\]
while for population~Q we used marginal probabilities 
\[
p^{(Q)} = (0.050,\, 0.005)
\] 
and odds ratio matrix 
\[
\begin{bmatrix}
- & 3 \\
3 & -
\end{bmatrix}.
\]
The values of these parameters were inspired by empirical data from a national upper secondary school leaving examination, as introduced in the real data example in Section~\ref{sec:RealDataExample}. The odds ratio matrices indicate that students from academic secondary schools ($C_1 = 1$) are less likely to take a repeated attempt ($C_2 = 0$), reflecting empirical patterns.

    \item We proceeded to generate a continuous variable, denoted as $C_3$, which is influenced by both binary variables ($C_1$ and $C_2$) and may represent a score from another test. For populations P and Q, the generation process was identical except for a possible constant shift. For scenarios in which the covariate test difficulty was assumed to be the same for both populations (scenarios 1--4 and 9--10; see Table~\ref{tab:scenarios}), the generation of $C_3$ follows the formula:
      \begin{align}\label{simulation - covariate with same difficulties}
         C_3 \sim \mathcal{N}(30, 17) + 10 \cdot C_1 + 25 \cdot C_2.
      \end{align}
For scenarios in which the covariate test difficulty differed between the populations (scenarios 5--8 and 11--12), the same formula was used for population P, while for population Q a constant shift was added:
      \begin{align}
         C_3 \sim \mathcal{N}(30, 17) + 10 \cdot C_1 + 25 \cdot C_2 + 10.
     \end{align}
     This shift represents a situation in which the covariate test administered to population Q is systematically easier.
    \item Finally, we computed the observed total scores based on the generated covariates. For population P, the total score, denoted as $X$, was determined by the following equation:
\begin{align}\label{total score X}
X \sim 20 \cdot C_1 + 5 \cdot C_2 + \alpha  \cdot C_3 + \beta + \epsilon.
\end{align}

Similarly, for population Q, the total score, denoted as $Y$, was first computed using the same equation:
\begin{align}\label{total score Y}
Y \sim 20 \cdot C_1 + 5 \cdot C_2 + \alpha \cdot C_3 + \beta + \epsilon.
\end{align}

Here, $\epsilon$ denotes a random error term following a normal distribution $\mathcal{N}(0,10)$. The $X$ and $Y$ values were then truncated to the test score range and rounded to the nearest integer.

The coefficients $\alpha$ and $\beta$ determined the role of the continuous covariate $C_3$ in the construction of the total scores. Specifically, $\alpha$ controlled the weight of $C_3$ relative to the binary covariates $C_1$ and $C_2$, while $\beta$ served as an intercept term ensuring that the resulting score distribution remained on a comparable scale across different simulation settings. By adjusting these two parameters, we were able to vary the strength of the association between $C_3$ and the total scores. 

We consider two basic options: 

\begin{enumerate}[label=\alph*)]
    \item \textit{Main simulation setting (strong relationship)}\\
   In this setting, we set $\alpha = 1$ and $\beta = 0$. With these choices, the value of $C_3$ was directly entered into the score equation without downscaling, resulting in a substantial contribution to the total score relative to the binary covariates. This configuration, therefore, represented a scenario in which the continuous covariate and the test scores were strongly related.  
    \item \textit{Alternative setting (weaker relationship)}\\
    In this setting, we set $\alpha = 0.5$ and $\beta = 30$. By halving the coefficient of $C_3$, its contribution to the total scores was reduced, meaning that $C_3$ played a smaller role in explaining score variation. To compensate for this reduced weight and to ensure that the scale of resulting scores remained comparable to the strong relationship scenario, we introduced a positive intercept of 30 points. Consequently, while $C_3$ still contributed to the scores, its influence was weaker, and the relationship between $C_3$ and the total scores was less pronounced. 
\end{enumerate}

Notably, in both these settings, the formulas for $X$ and $Y$ are identical, which indicates that the tests represented by these total scores were assumed to be of comparable difficulty in expectation. The true equating transformation was derived analytically from the known data-generating model, consistent with the model-based approach to equating criteria in simulation studies \parencite{gonzalez2017applying, wiberg2025generalized}. In Scenarios 1--8, X and Y were generated from identical model equations (Equations~\eqref{total score X} and \eqref{total score Y}), so the true transformation reduces to the identity function. While the identity transformation may not provide a realistic benchmark for bias evaluation when populations or test forms differ in real-world settings \parencite{Wiberg2025BiasEquating}, in this simulation, it serves as a controlled reference point for interpreting bias in scenarios where $X$ and $Y$ are generated from the same model.

To further investigate the performance of the approach under a more challenging setting, we also considered scenarios in which the overall test difficulty differed between populations. Specifically, for population $Q$, the total scores were adjusted as $Y' = 0.9\cdot Y + 5$, while the total scores for population $P$ remained unchanged. This modification allowed us to examine the behavior of the approach when differences in test difficulty are present (Scenarios 9--12). In these scenarios, the true equating transformation linking $X$ to the $Y'$ scale follows directly from this known linear relationship between the two score distributions. 
\end{enumerate}

To provide a clear overview of the simulation design, Table~\ref{tab:scenarios} summarizes all scenarios considered in this study, including the covariate test difficulty conditions, the strength of the relationship between the continuous covariate $C_3$ and the total scores, the corresponding coefficients $\alpha$ and $\beta$, the adjustments applied to total scores to reflect differences in overall test difficulty and the sample sizes used for each scenario.

\begin{table}[ht]
\centering
\begin{tabular}{crrrrrr}
\hline
Scenario & \multicolumn{1}{c}{Relationship} & \multicolumn{1}{c}{Shift in $C_3$} & \multicolumn{1}{c}{$Y'$ transformation} & \multicolumn{1}{c}{$\alpha$} & \multicolumn{1}{c}{$\beta$} & \multicolumn{1}{c}{Sample size} \\
\hline
1        & Strong & $0$ & $Y$    & $1$   & $0$  & 5,000  \\
2       & Strong & $0$   & $Y$  & $1$   & $0$  & 50,000 \\
3        & Weak   & $0$  & $Y$   & $0.5$ & $30$ & 5,000  \\
4       & Weak   & $0$  & $Y$   & $0.5$ & $30$ & 50,000 \\
5   & Strong & $+10$  & $Y$ & $1$   & $0$  & 5,000  \\
6   & Strong & $+10$  & $Y$ & $1$   & $0$  & 50,000 \\
7   & Weak   & $+10$ & $Y$  & $0.5$ & $30$ & 5,000  \\
8   & Weak   & $+10$  & $Y$ & $0.5$ & $30$ & 50,000 \\
9  & Strong & 0   & 0.9$\cdot Y + 5$  & 1   & 0   & 5,000   \\
10 & Strong & 0   & 0.9$\cdot Y + 5$ & 1   & 0   & 50,000   \\
11 & Strong & +10 & 0.9$\cdot Y + 5$  & 1   & 0  & 5,000    \\
12 & Strong & +10 & 0.9$\cdot Y + 5$  & 1   & 0  & 50,000   \\
\hline
\end{tabular}
\caption{Overview of simulation scenarios including the strength of the $C_3$ covariate and total score relationship, shifts in $C_3$, adjustments in total score $Y$ for population Q, and sample sizes.}
\label{tab:scenarios}
\end{table}

Data generation and equating analyses were conducted in R version 4.3.2 \parencite{R} with the \texttt{kequate} package \parencite{Andersson2013kequate}. We applied \gls{gke} with the \gls{nec} design as described in Section~\ref{sec2.1:kernel covariate equating}, using raw (non-equated) covariate values to equate the total scores $X$ and $Y$ directly. When the covariate $C_3$ differed in difficulty across groups (scenarios 5--8 and 11--12, see Table~\ref{tab:scenarios}), sequential \gls{gke} was also applied. In sequential \gls{gke}, the covariate $C_3$ was first equated across groups using \gls{gke} with attended school type ($C_1$) and exam attempt ($C_2$) included as covariates, after which the equated covariate values were incorporated into the subsequent equating of the total scores $X$ and $Y$. Covariate $C_3$ was always categorized into five fixed categories with thresholds at 50, 60, 70, 80, and 100 score points. In all cases, presmoothing of the observed score distributions was performed using log-linear Poisson regression models. Specifically, polynomial terms of the score up to degree 6, main effects of the covariates, and their interactions with the score were included. The models were selected using forward--backward stepwise selection based on the statistical significance of the terms, while preserving model hierarchy. Details of the selected presmoothing models, including AIC and BIC values, are provided in the Supplemental Material. The continuization step in \gls{gke} was achieved using a Gaussian kernel, with the bandwidth automatically selected in each replication using a penalty criterion \parencite{vonDavier2004kernel} and the KPEN parameter set to 1. At each score point,  the resulting equated values were compared to the true equating transformation using the evaluation measures defined in Section~\ref{sec2.1:kernel covariate equating}. In the simulation, $R=100$ replications were used. Bias, standard errors of equating, and \gls{rmse} were computed at each score point. In addition, \gls{ediff} were used to compare the \gls{gke} and sequential \gls{gke} transformations, and the resulting differences were interpreted relative to the \gls{dtm} criterion. 

\subsection{Results of the simulation study}

\begin{figure}[t]
    \centering
    \begin{subfigure}[b]{0.32\textwidth}
        \centering
        \includegraphics[width=\textwidth]{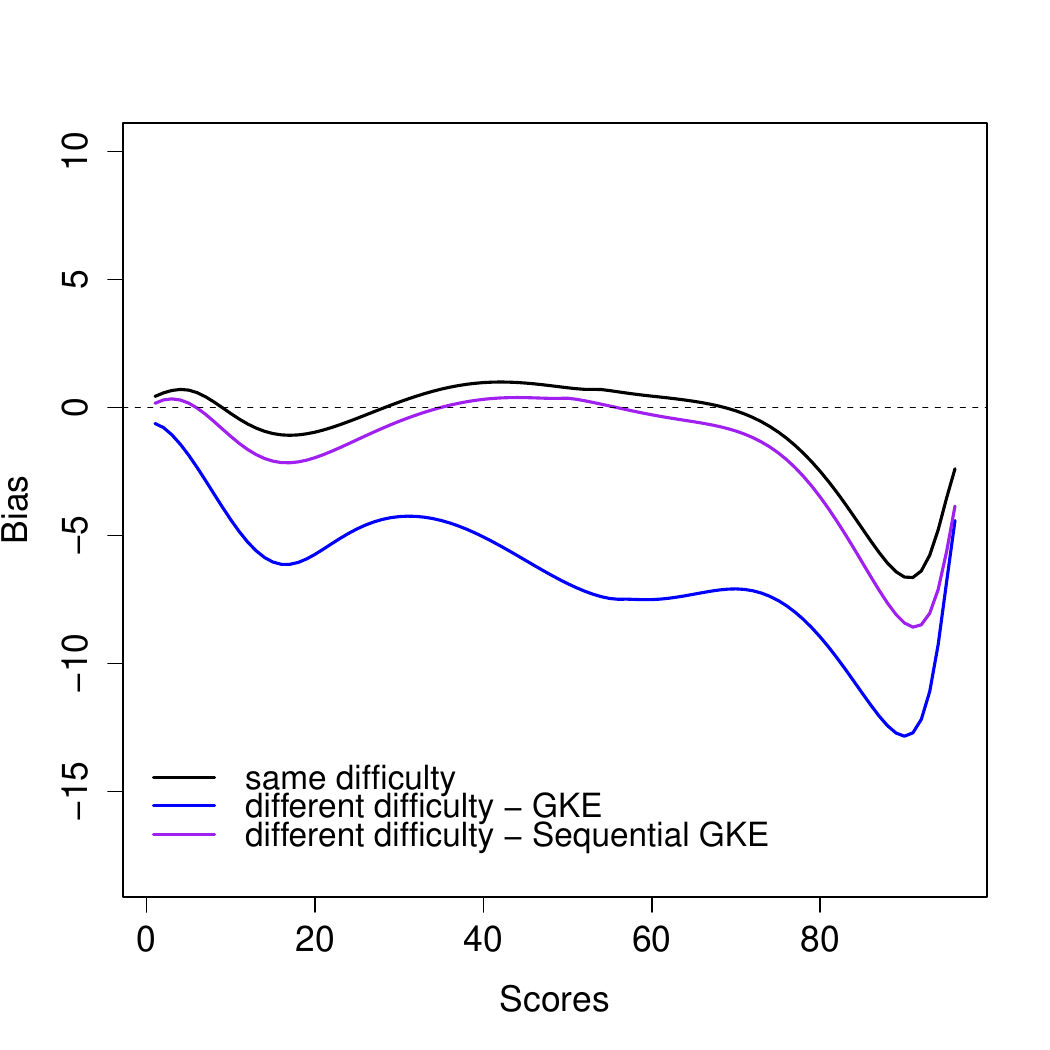}
        \caption{Bias}
    \end{subfigure}
    \hfill
    \begin{subfigure}[b]{0.32\textwidth}
        \centering
        \includegraphics[width=\textwidth]{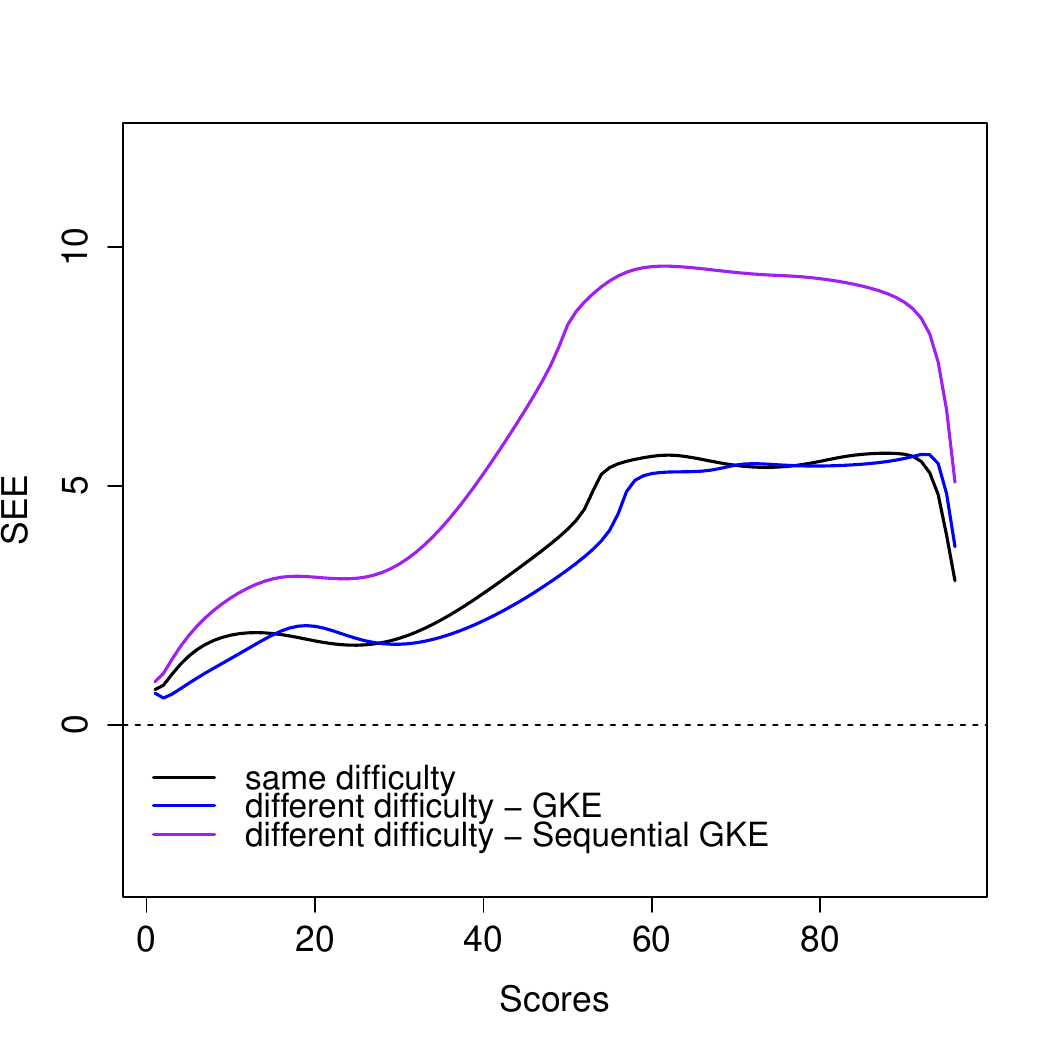}
        \caption{SEE}
    \end{subfigure}
    \hfill
    \begin{subfigure}[b]{0.32\textwidth}
        \centering
        \includegraphics[width=\textwidth]{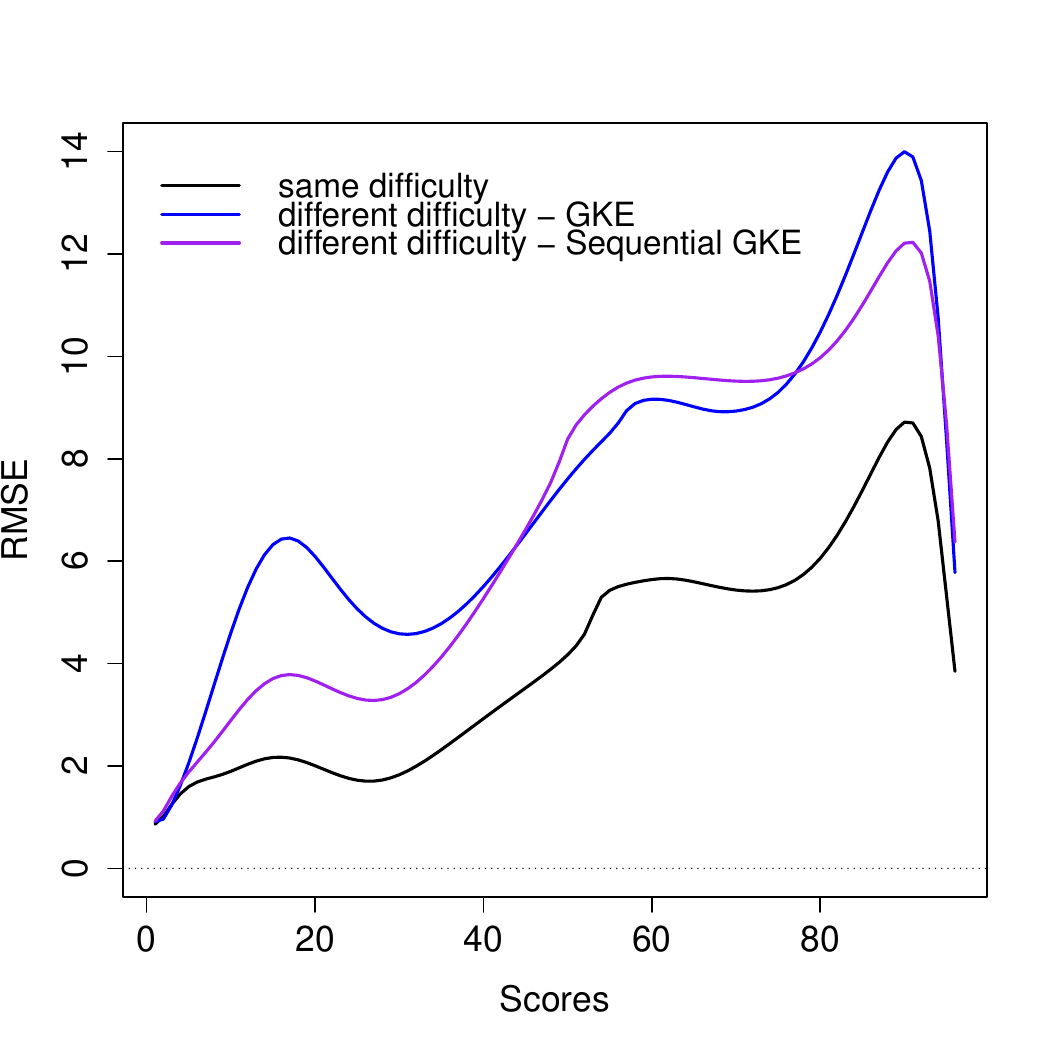}
        \caption{RMSE}
    \end{subfigure}
    \caption{Bias (a), \gls{see} (b), \gls{rmse} (c) in models where test score covariate was generated as coming from test with the same difficulty (solid black line), as coming from test with a different difficulty (dashed blue line) and as coming from test with a different difficulty but equated before including to the model (dot-dashed purple line), for sample size 5,000 test takers. Total scores $X$ and $Y$ were generated assuming a strong relationship between them and the test score covariate, following the equations \eqref{total score X} and \eqref{total score Y} with parameters $\alpha = 1$ and $\beta = 0$, see Scenarios 1 and 5 in Table~\ref{tab:scenarios}.}
    \label{fig:Simulation5000strong}
\end{figure}

\begin{figure}[t]
    \centering
    \begin{subfigure}[b]{0.32\textwidth}
        \centering
        \includegraphics[width=\textwidth]{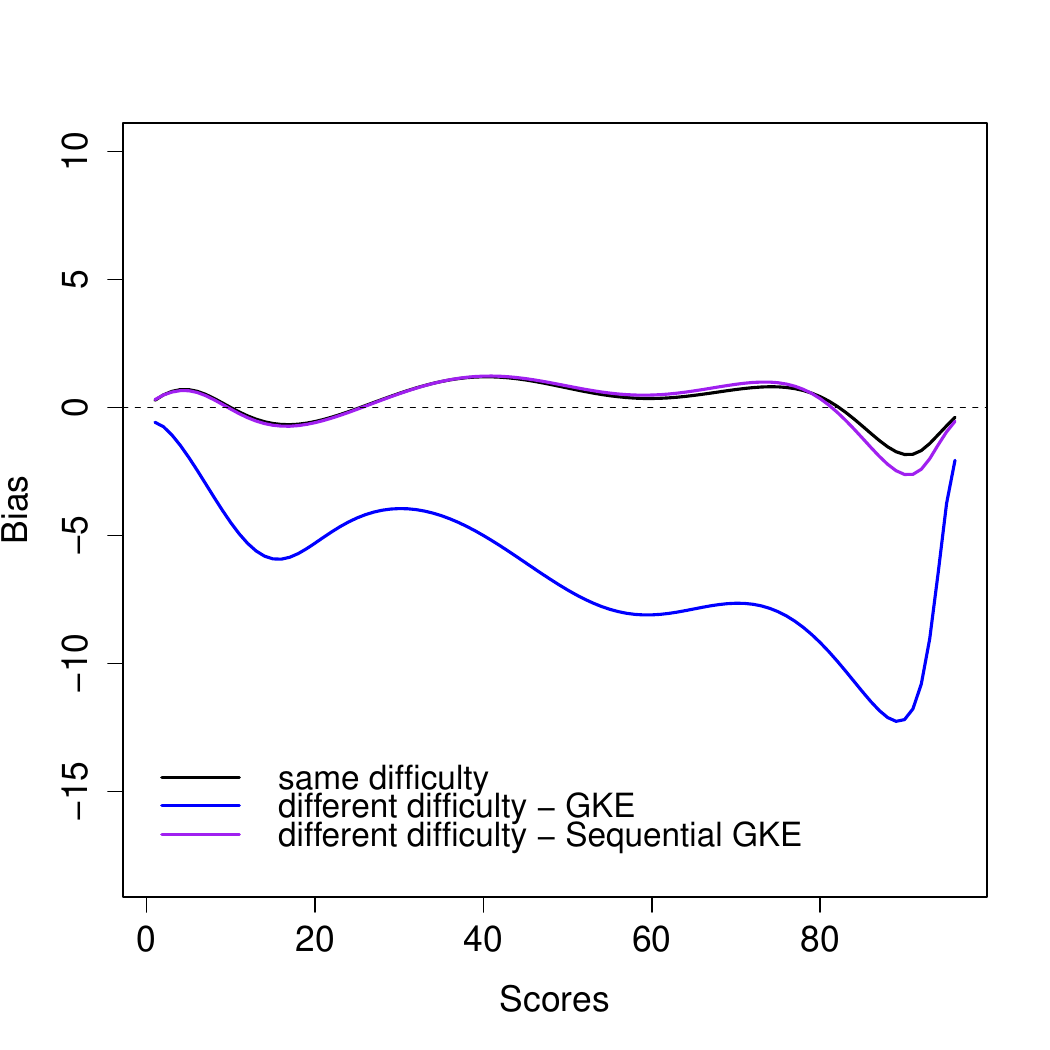}
        \caption{Bias}
    \end{subfigure}
    \hfill
    \begin{subfigure}[b]{0.32\textwidth}
        \centering
        \includegraphics[width=\textwidth]{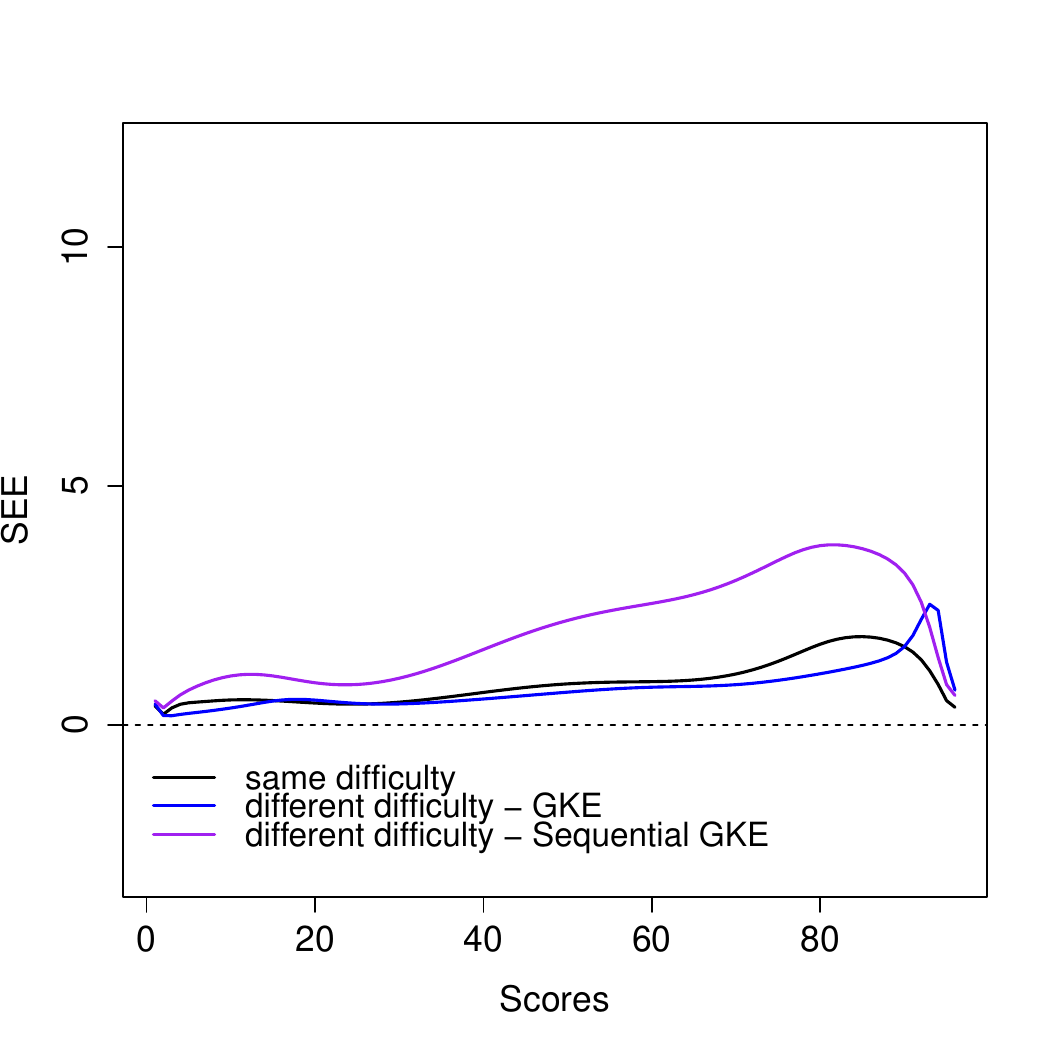}
        \caption{SEE}
    \end{subfigure}
    \hfill
    \begin{subfigure}[b]{0.32\textwidth}
        \centering
        \includegraphics[width=\textwidth]{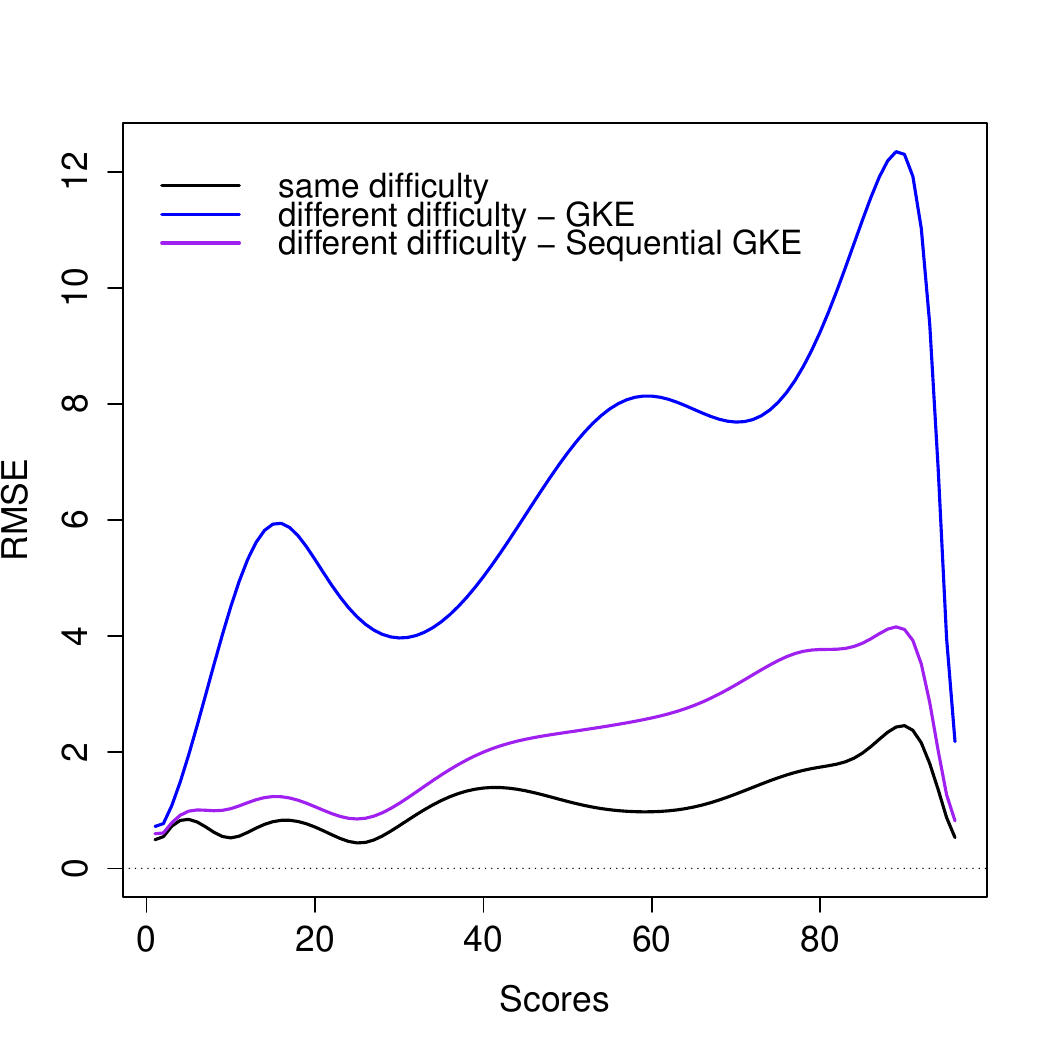}
        \caption{RMSE}
    \end{subfigure}
    \caption{Bias (a), \gls{see} (b), \gls{rmse} (c) in models where test score covariate was generated as coming from test with the same difficulty (solid black line), as coming from test with a different difficulty (dashed blue line) and as coming from test with a different difficulty but equated before including to the model (dot-dashed purple line), for sample size 50,000 test takers. Total scores $X$ and $Y$ were generated assuming a strong relationship between them and the test score covariate, following the equations \eqref{total score X} and \eqref{total score Y} with parameters $\alpha = 1$ and $\beta = 0$, see Scenarios 2 and 6 in Table~\ref{tab:scenarios}.}
    \label{fig:Simulation50000strong}
\end{figure}

\begin{figure}[t]
    \centering
    \begin{subfigure}[b]{0.32\textwidth}
        \centering
        \includegraphics[width=\textwidth]{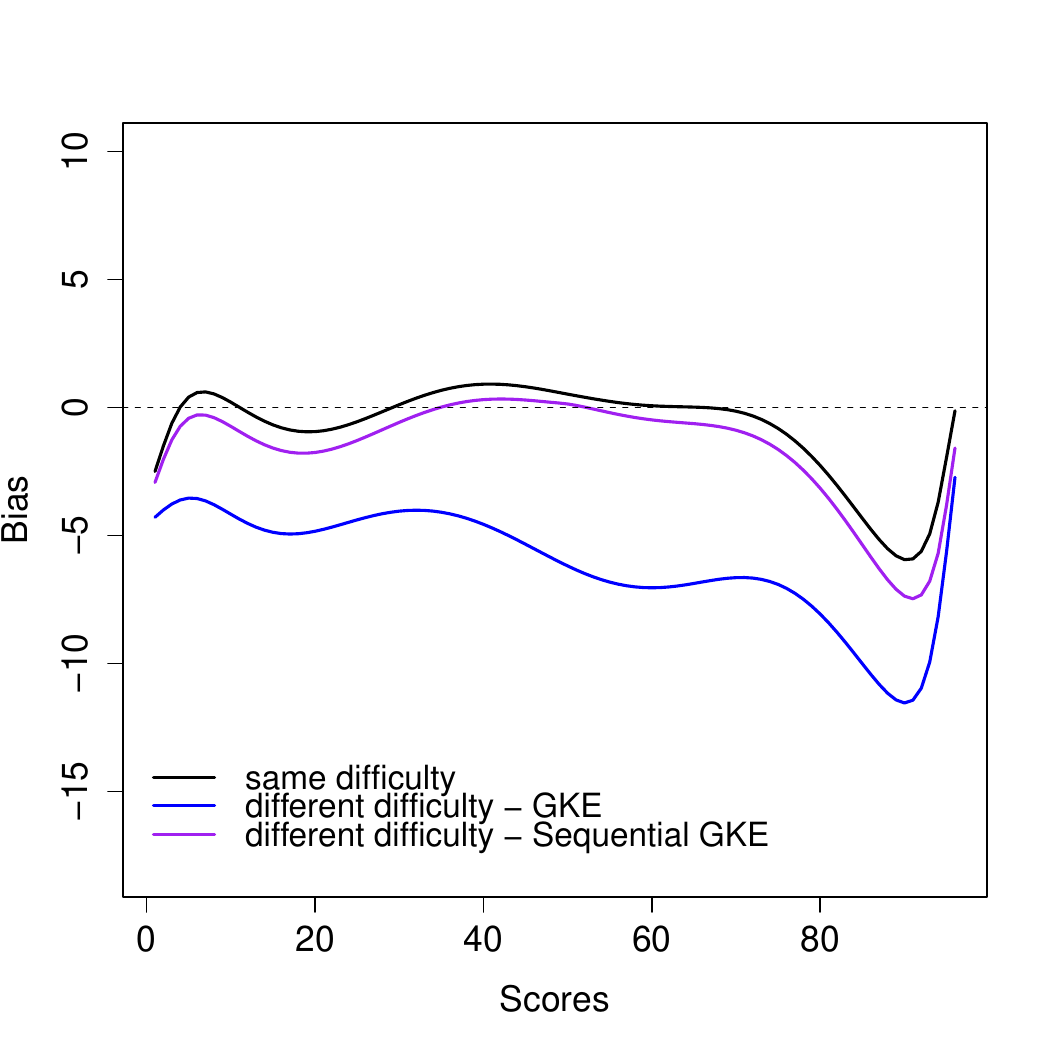}
        \caption{Bias}
    \end{subfigure}
    \hfill
    \begin{subfigure}[b]{0.32\textwidth}
        \centering
        \includegraphics[width=\textwidth]{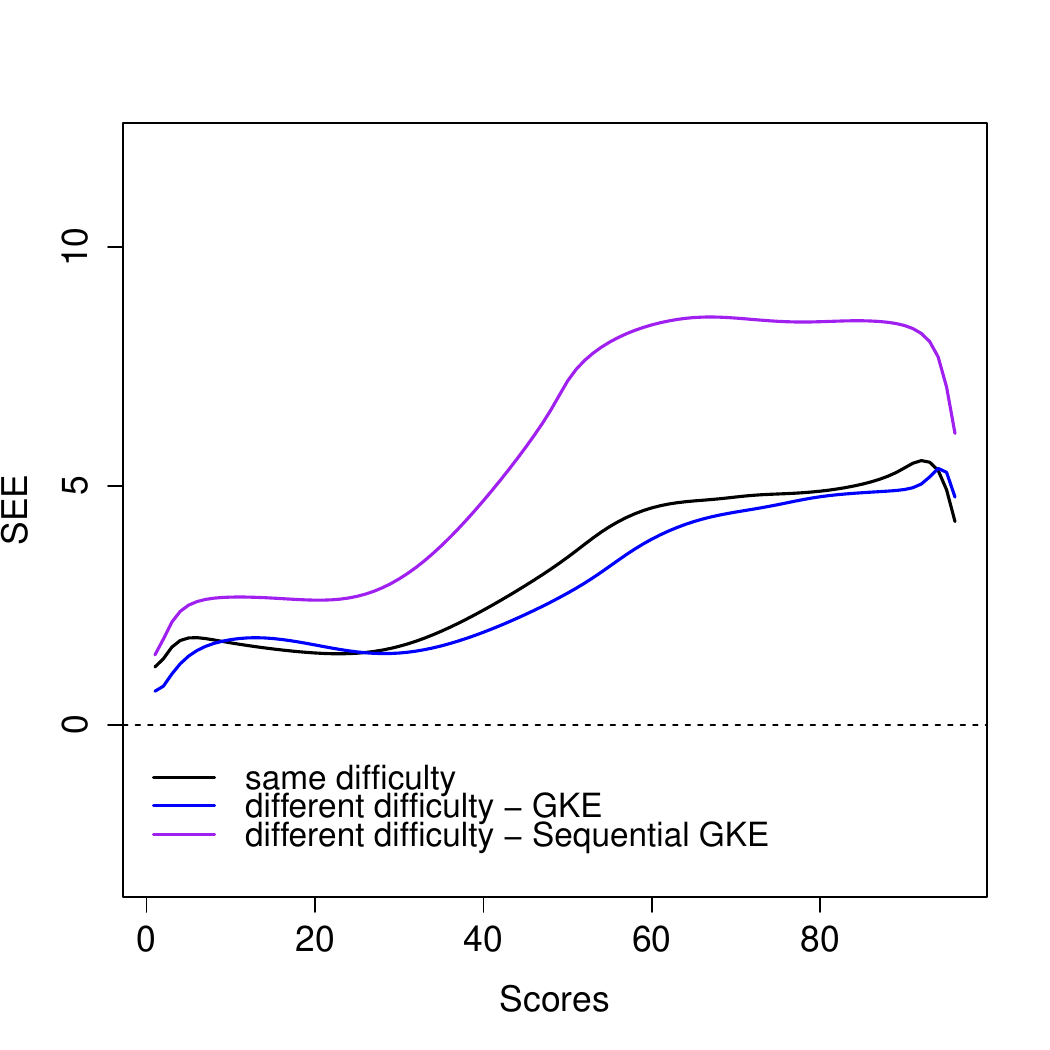}
        \caption{SEE}
    \end{subfigure}
    \hfill
    \begin{subfigure}[b]{0.32\textwidth}
        \centering
        \includegraphics[width=\textwidth]{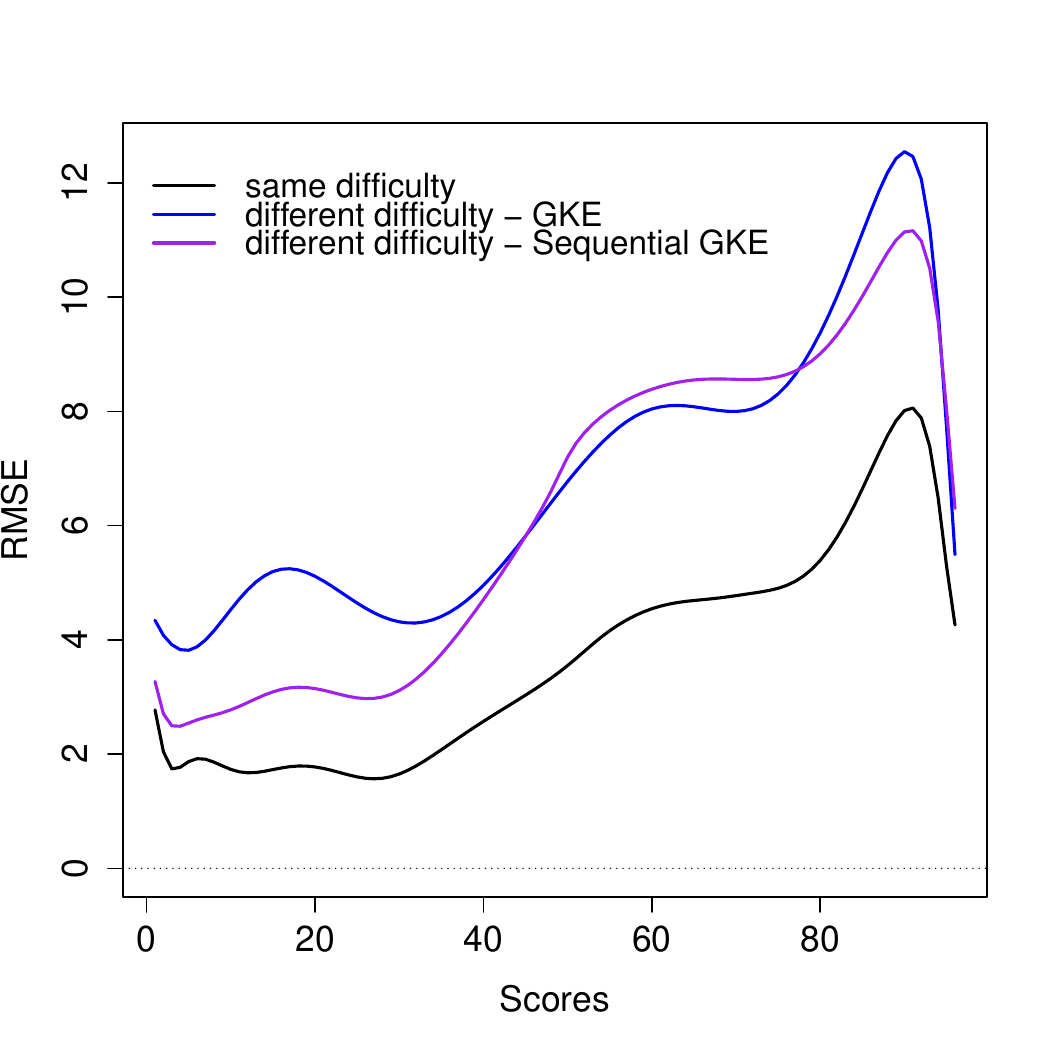}
        \caption{RMSE}
    \end{subfigure}
    \caption{Bias (a), \gls{see} (b), \gls{rmse} (c) in models where the test score covariate was generated as coming from a test with the same difficulty (solid black line), from a test with a different difficulty (dashed blue line), and from a test with a different difficulty but equated before inclusion in the model (dot-dashed purple line), for sample sizes of 5,000 test takers. Total scores $X$ and $Y$ were generated assuming a strong relationship with the test score covariate, following the equations \eqref{total score X} and \eqref{total score Y} with parameters $\alpha = 1$ and $\beta = 0$, with $Y'$ transformation as specified in Scenarios 9 and 11 of Table~\ref{tab:scenarios}.}
    \label{fig:Simulation5000strongTr}
\end{figure}

\begin{figure}[t]
    \centering
    \begin{subfigure}[b]{0.32\textwidth}
        \centering
        \includegraphics[width=\textwidth]{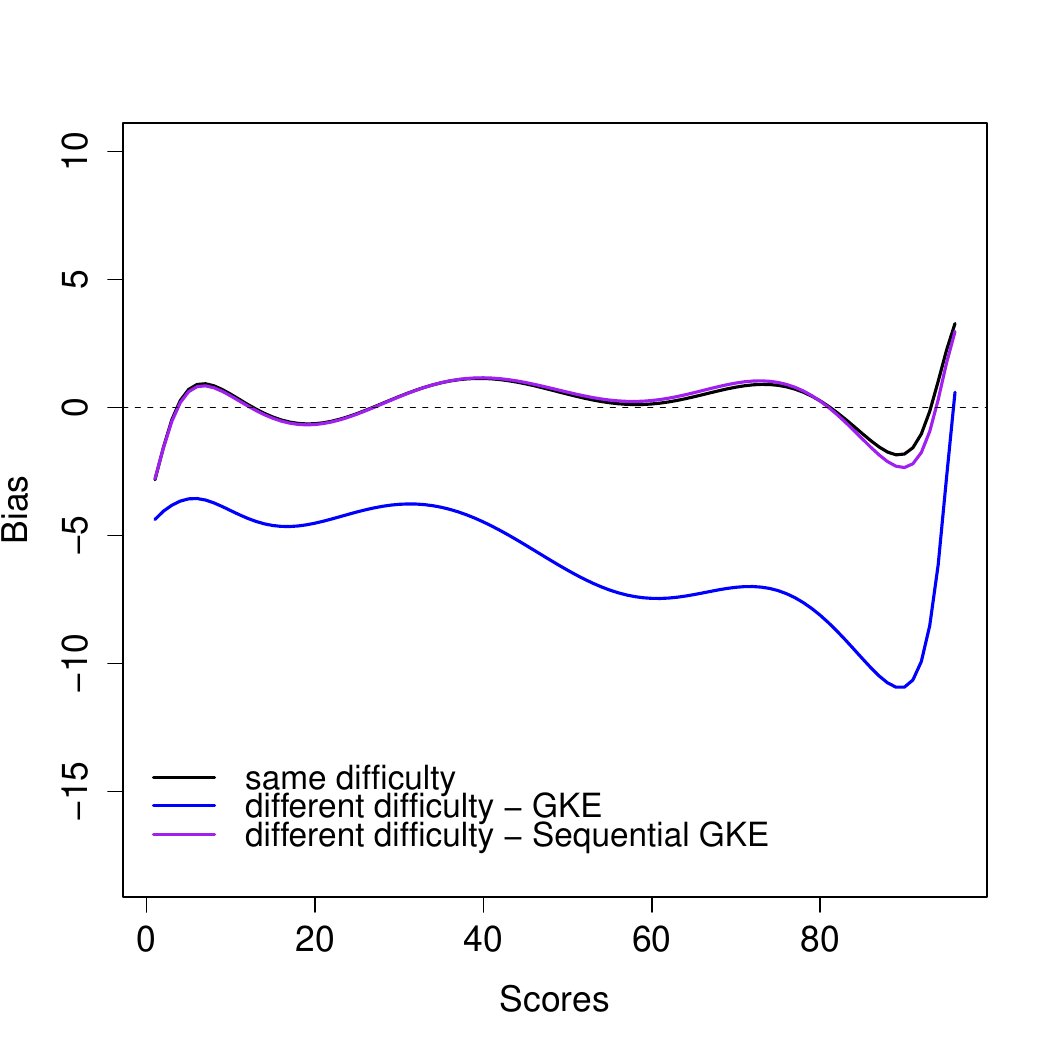}
        \caption{Bias}
    \end{subfigure}
    \hfill
    \begin{subfigure}[b]{0.32\textwidth}
        \centering
        \includegraphics[width=\textwidth]{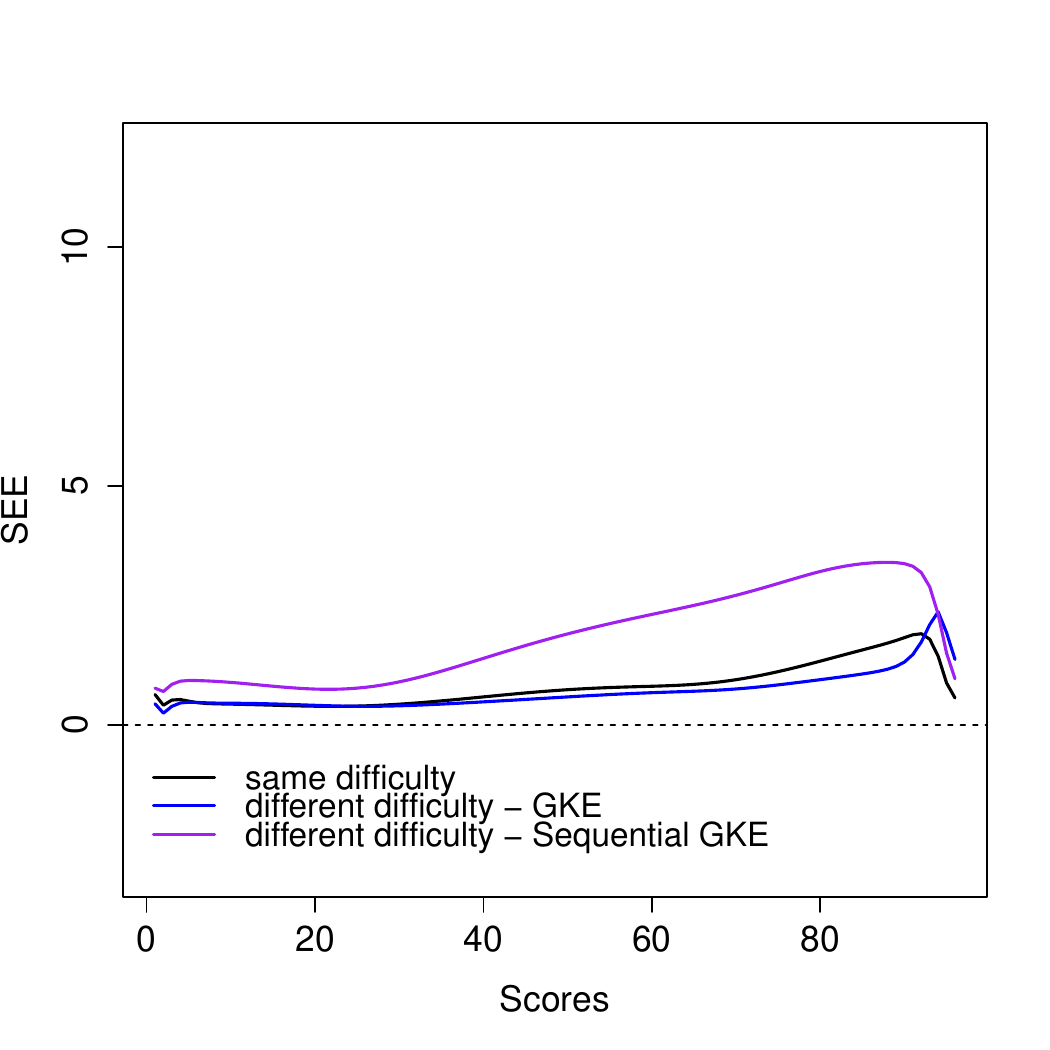}
        \caption{SEE}
    \end{subfigure}
    \hfill
    \begin{subfigure}[b]{0.32\textwidth}
        \centering
        \includegraphics[width=\textwidth]{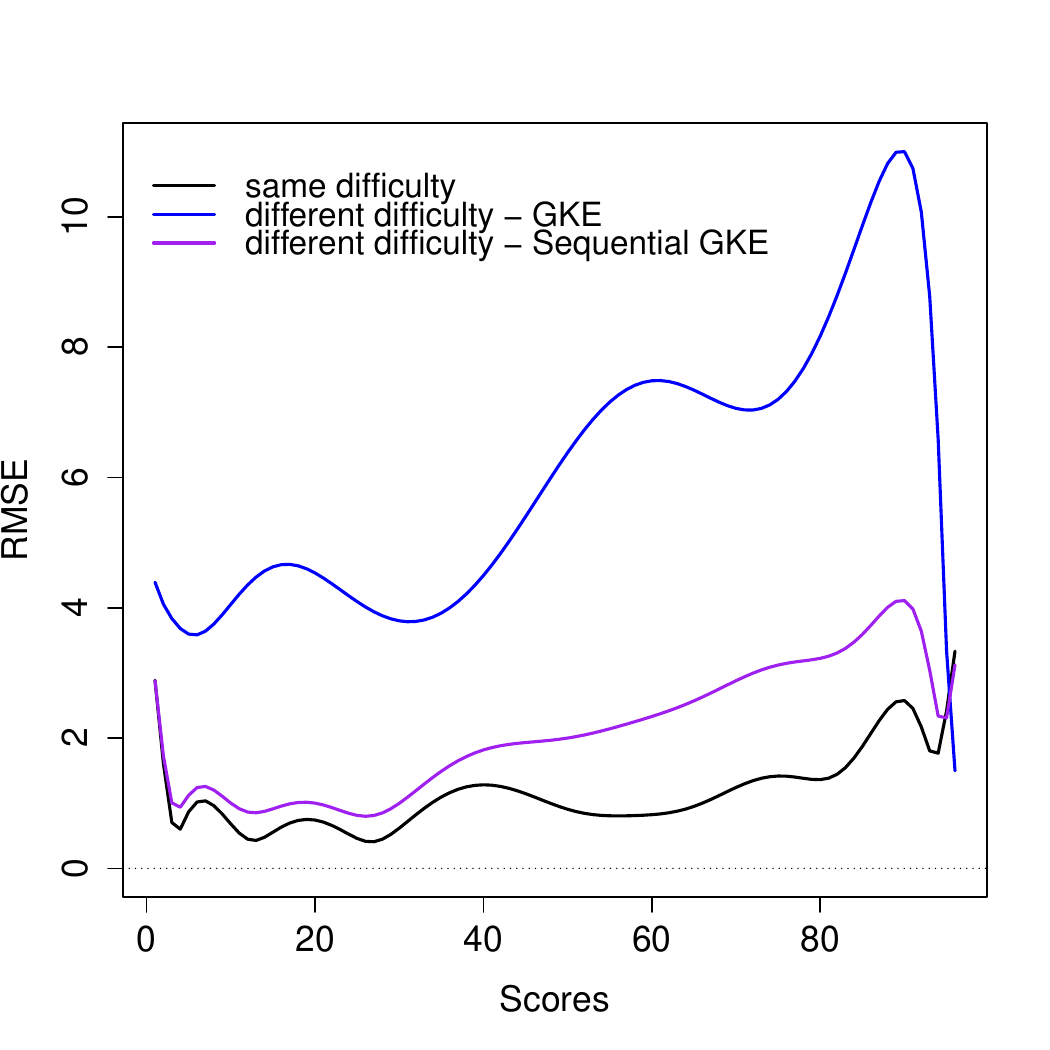}
        \caption{RMSE}
    \end{subfigure}
    \caption{Bias (a), \gls{see} (b), \gls{rmse} (c) in models where the test score covariate was generated as coming from a test with the same difficulty (solid black line), from a test with a different difficulty (dashed blue line), and from a test with a different difficulty but equated before inclusion in the model (dot-dashed purple line), for sample sizes of 50,000 test takers. Total scores $X$ and $Y$ were generated assuming a strong relationship with the test score covariate, following the equations \eqref{total score X} and \eqref{total score Y} with parameters $\alpha = 1$ and $\beta = 0$, with $Y'$ transformation as specified in Scenarios 10 and 12 of Table~\ref{tab:scenarios}.}
    \label{fig:Simulation50000strongTr}
\end{figure}

The results demonstrated that with a strong relationship between the total test score and the continuous covariate $C_3$, when the assumption of population-invariant conditional distributions was met, the bias in equated scores was minimal, especially with a large sample size (black line in Figures~\ref{fig:Simulation5000strong} and~\ref{fig:Simulation50000strong}). However, when the assumption was violated by introducing different test difficulties, the bias in equated scores increased remarkably, even with a large sample size (blue line in Figures~\ref{fig:Simulation5000strong} and~\ref{fig:Simulation50000strong}). %This was particularly evident when the relationship between the total score and the continuous variable $C_3$ was strong, as illustrated in Figure~\ref{main scenario results}. 
Equating the covariate distribution before performing \gls{gke} with the primary test scores significantly increased the accuracy of the main equating process (purple line in Figures~\ref{fig:Simulation5000strong} and~\ref{fig:Simulation50000strong}). Similar results were observed for scenarios with equal as well as different difficulty levels of the primary test forms (see Figures \ref{fig:Simulation5000strongTr} and \ref{fig:Simulation50000strongTr}). While equating the covariate led to an increase in \gls{see}, the overall \gls{rmse} was improved, particularly for larger sample sizes, indicating a favorable trade-off between precision and accuracy.

\begin{figure}[t]
    \centering
    \begin{subfigure}[b]{0.32\textwidth}
        \centering
        \includegraphics[width=\textwidth]{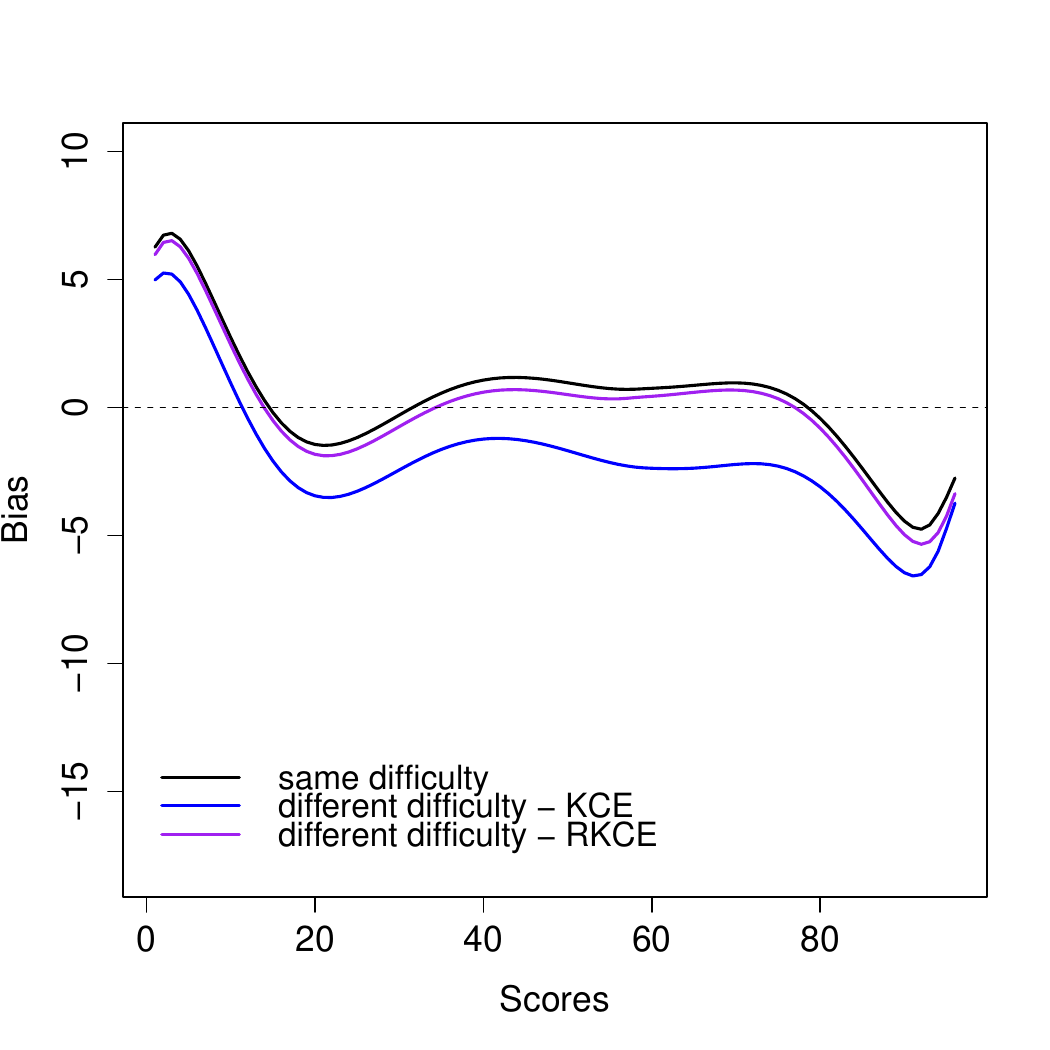}
        \caption{Bias}
    \end{subfigure}
    \hfill
    \begin{subfigure}[b]{0.32\textwidth}
        \centering
        \includegraphics[width=\textwidth]{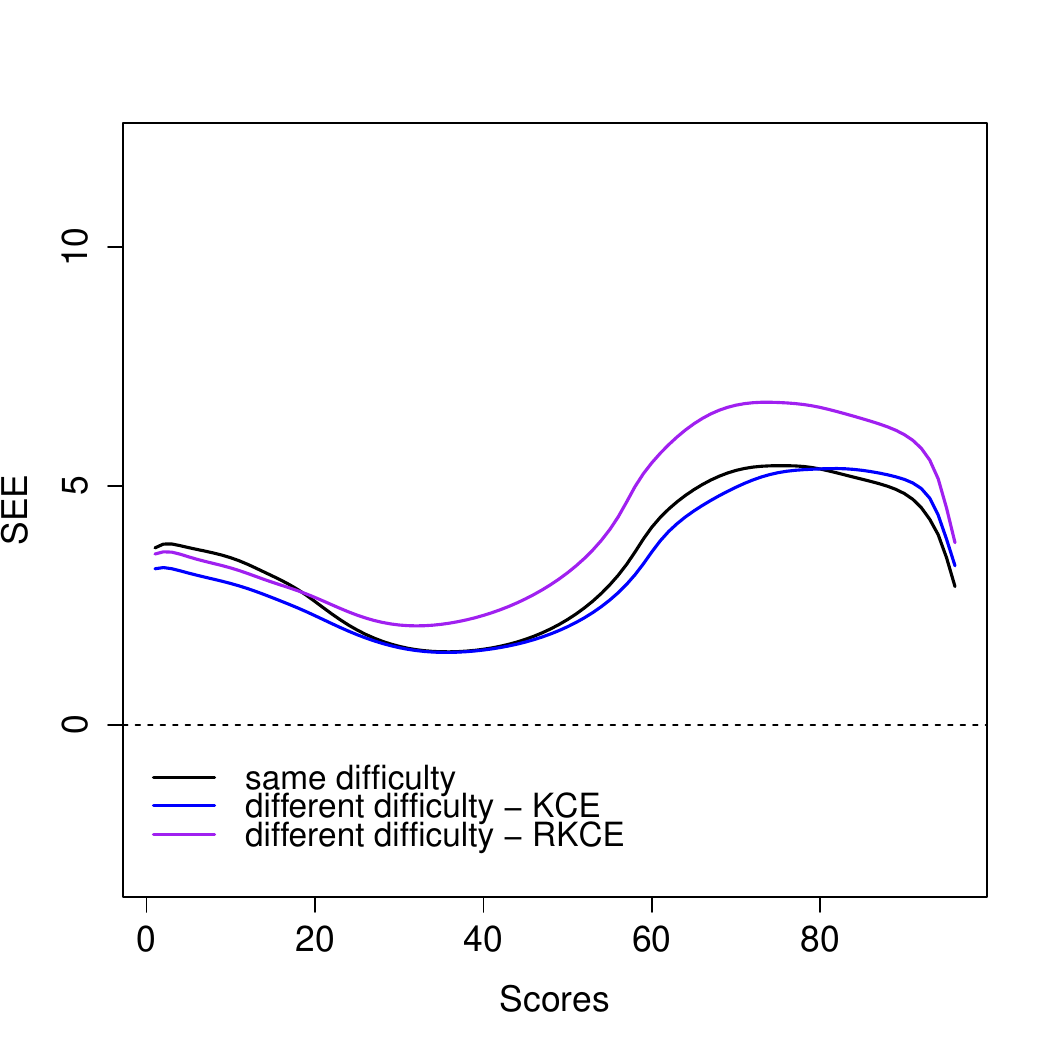}
        \caption{SEE}
    \end{subfigure}
    \hfill
    \begin{subfigure}[b]{0.32\textwidth}
        \centering
        \includegraphics[width=\textwidth]{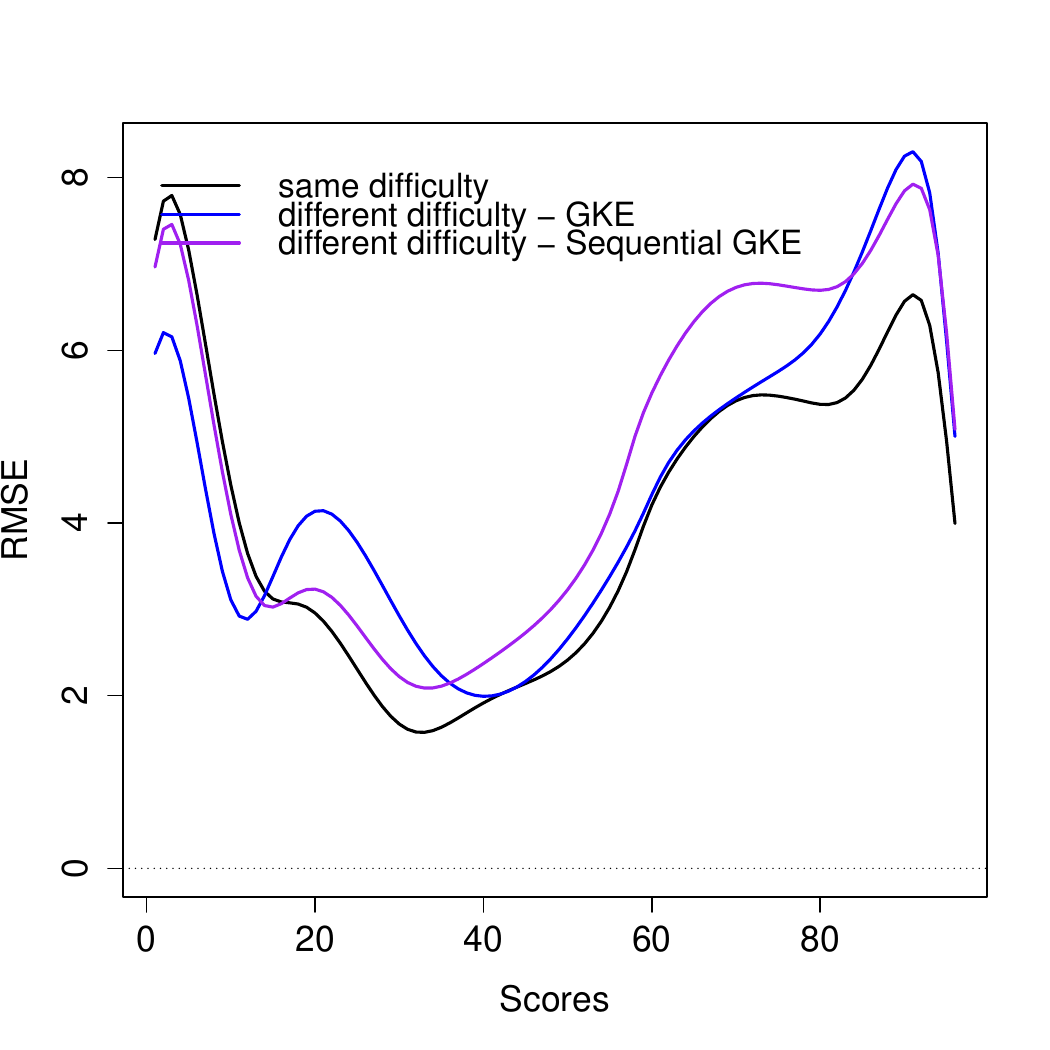}
        \caption{RMSE}
    \end{subfigure}
    \caption{Bias (a), \gls{see} (b), \gls{rmse} (c) in models where test score covariate was generated as coming from test with the same difficulty (solid black line), as coming from test with a different difficulty (dashed blue line) and as coming from test with a different difficulty but equated before including to the model (dot-dashed purple line), for sample sizes 5,000 test takers. Total scores $X$ and $Y$ were generated assuming a weaker relationship between them and the test score covariate, following the equations \eqref{total score X} and \eqref{total score Y}  with parameters $\alpha = 0.5$ and $\beta = 30$, see Scenarios 3 and 7 in Table~\ref{tab:scenarios}.}
    \label{fig:Simulation5000weak}
\end{figure}

\begin{figure}[t]
    \centering
    \begin{subfigure}[b]{0.32\textwidth}
        \centering
        \includegraphics[width=\textwidth]{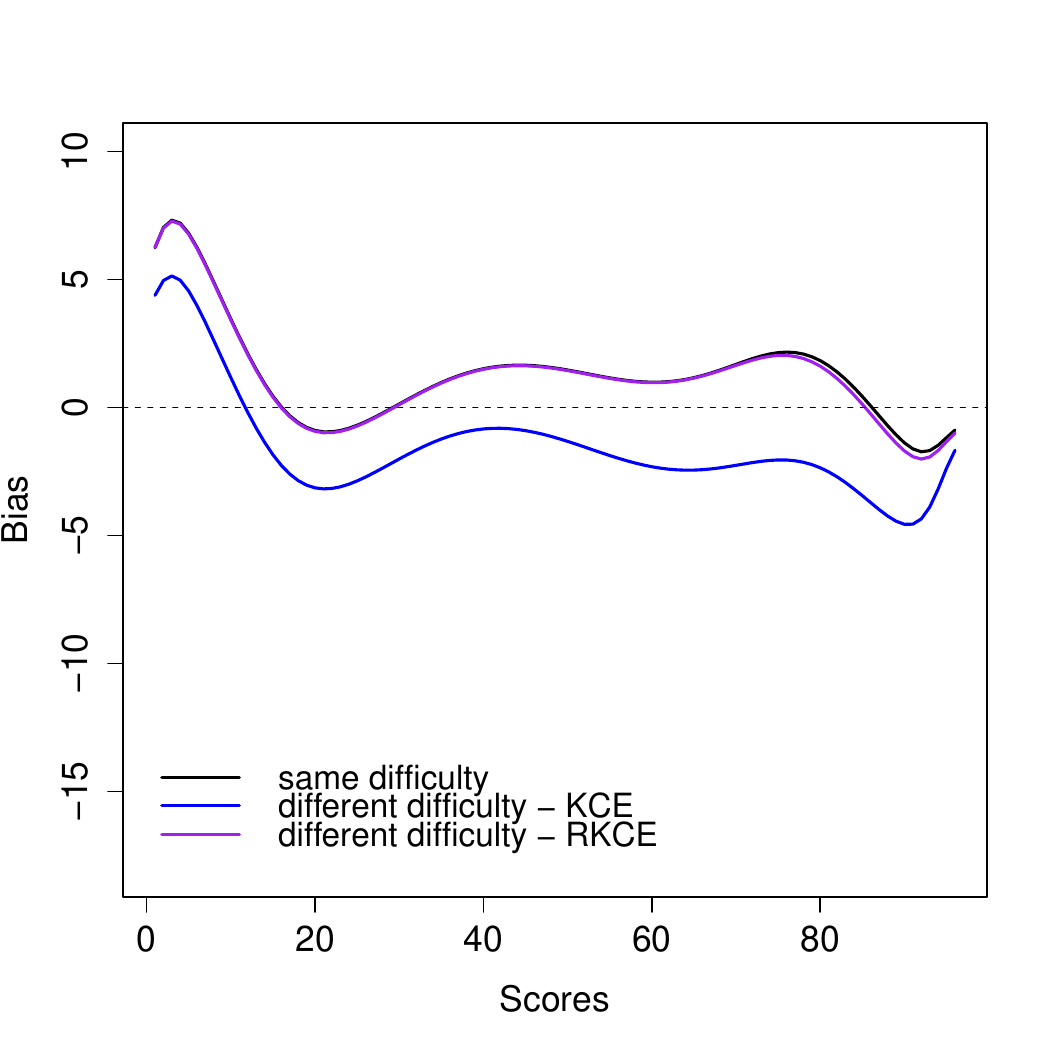}
        \caption{Bias}
    \end{subfigure}
    \hfill
    \begin{subfigure}[b]{0.32\textwidth}
        \centering
        \includegraphics[width=\textwidth]{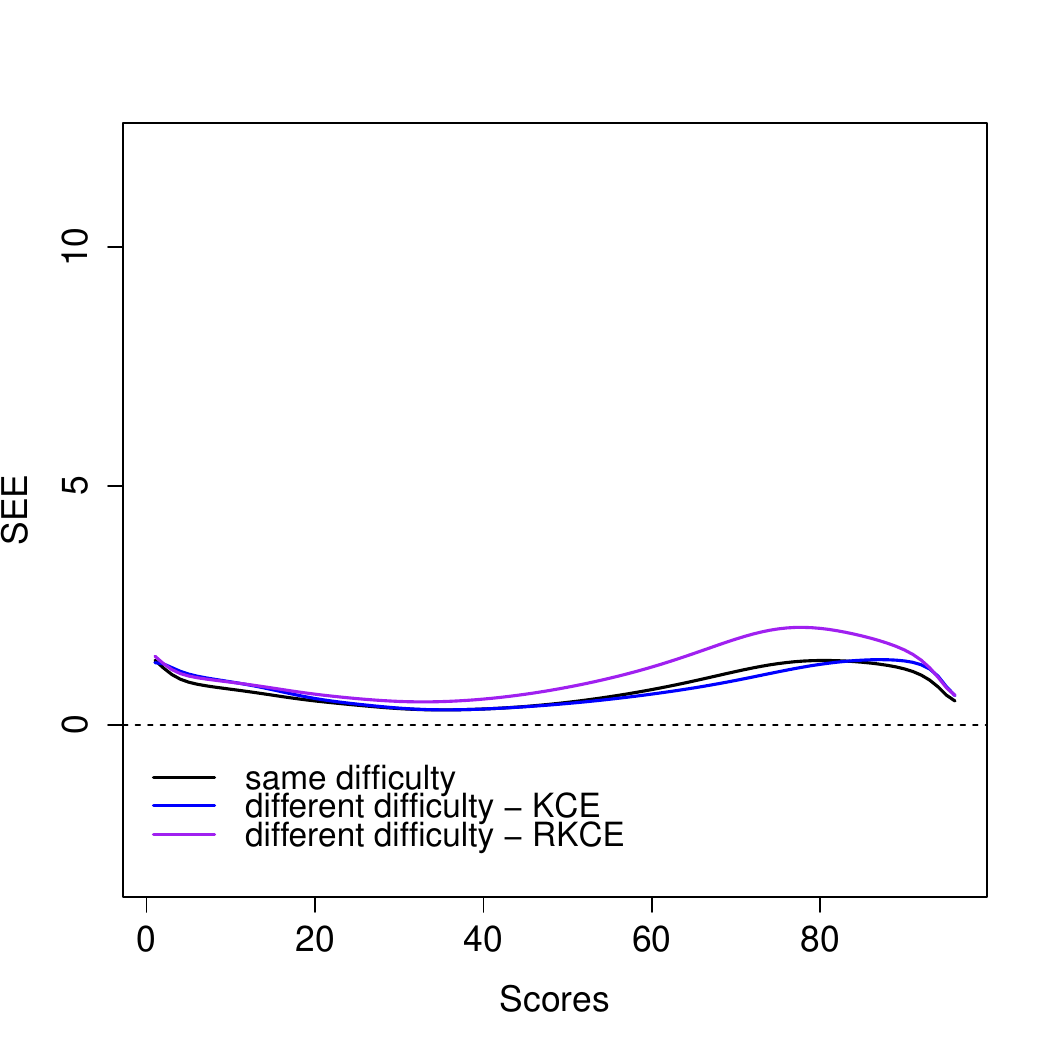}
        \caption{SEE}
    \end{subfigure}
    \hfill
    \begin{subfigure}[b]{0.32\textwidth}
        \includegraphics[width=\textwidth]{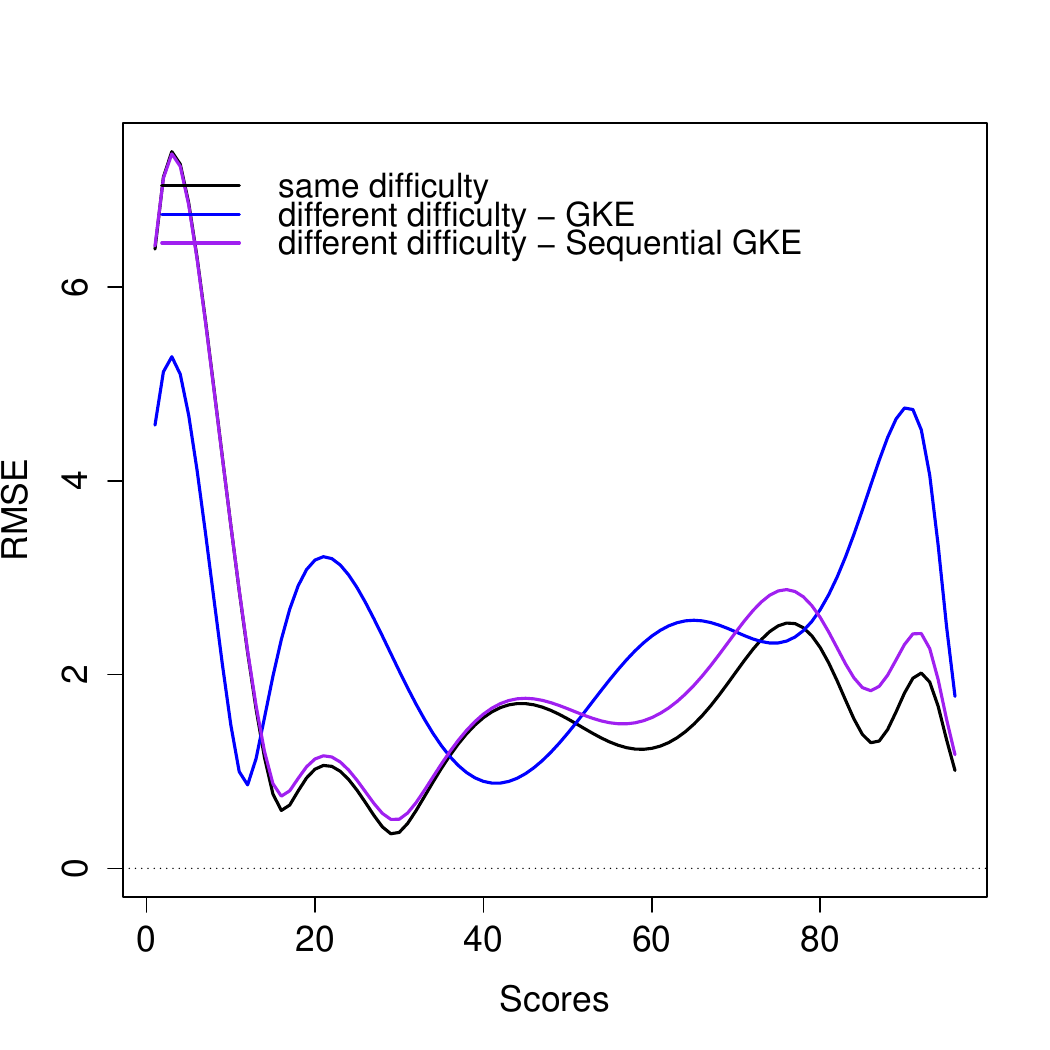}
        \caption{RMSE}
    \end{subfigure}
    \caption{Bias (a), \gls{see} (b), \gls{rmse} (c) in models where test score covariate was generated as coming from test with the same difficulty (solid black line), as coming from test with a different difficulty (dashed blue line) and as coming from test with a different difficulty but equated before including to the model (dot-dashed purple line), for sample sizes 50,000 test takers. Total scores $X$ and $Y$ were generated assuming a weaker relationship between them and the test score covariate, following the equations \eqref{total score X} and \eqref{total score Y}  with parameters $\alpha = 0.5$ and $\beta = 30$, see Scenarios 4 and 8 in Table~\ref{tab:scenarios}.}
    \label{fig:Simulation50000weak}
\end{figure}

When the relationship between the continuous covariate and the test scores was weaker, the bias introduced by measuring the continuous covariate on shifted scales was less pronounced, since the $C_3$ variable had less influence on the total score, as illustrated in Figures~\ref{fig:Simulation5000weak} and~\ref{fig:Simulation50000weak}.  
However, equating the covariate helped reduce bias in the equating process even in this case without substantial \gls{see} increase.

The \gls{ediff} values were relatively large across all scenarios, which is expected given the substantial differences introduced in the covariate $C_3$ between populations. Consequently, the \gls{dtm} threshold of one half of a score point was exceeded for most score points in all scenarios. Mean \gls{ediff} across score points was 5.32 in Scenario 5, 6.66 in Scenario 6, 2.13 in Scenario 7, 2.76 in Scenario 8, 4.94 in Scenario 11, and 6.14 in Scenario 12.

%------------------------------------------------------------
\section{Real data example}\label{sec:RealDataExample}
To illustrate how the proposed two-step approach can be applied to address challenges in real-world educational settings, we perform both \gls{gke} and sequential \gls{gke} using data from the national upper secondary school leaving exam \parencite{MartinkovaPotuznikovaNetik_chapter}. The exam consists of four school subjects, from which only the \gls{nl} test is mandatory for all examinees. The students then choose between a Mathematics or a Foreign Language (mostly English) test and two oral or practical exams from a wider subject selection offered at the school level. Most students take the exam in the spring. Those who were not admitted to the spring term (i.e., did not complete the final school year, typically due to a high absence rate), missed or failed the exam, have another possibility in the fall term if they apply for it and complete the school year in time; otherwise, they will take the exam next spring. Students taking the exam in the spring and fall terms show notable differences in their profiles and exam performance, with fall examinees scoring significantly lower. 

None of the national tests includes anchor items that would enable test equating using the \gls{neat} design. Nevertheless, evidence of test score comparability between the two examination terms within the same year is essential for the fairness of the assessment, as a pass grade is a prerequisite for admission to tertiary education. Additionally, the comparability of test scores from the same subject over multiple years is desirable, allowing for the monitoring of trends in student achievement. While the spring-term \gls{nl} test is taken by a virtually complete cohort of upper secondary school students with comparable ability distributions across years, the comparability of scores in non-mandatory tests over multiple years poses an additional challenge, as the composition of students who opt for each non-mandatory subject (including English and Mathematics) varies from year to year, depending on changing preferences and possible changes in organizational arrangements that may influence students’ decisions about which subject they select. For example, although the national \gls{el} test was developed according to the same blueprint and item specifications across the years, the writing component of the exam (essay) was shifted from the national to the school level, leading to a decrease in \gls{el} preferences among academic school students, likely due to the anticipated higher difficulty of the writing part when administered and scored locally. As a result, contrary to the mandatory \gls{nl} test, any observed \gls{el} test score differences confound differences due to varying test difficulty and varying test-takers' ability even across the spring terms, calling for a \gls{nec} design when an anchor test is not available.      

\subsection{Data and analytical strategy}

We use data from the \gls{el} tests administered in the spring and fall terms from 2017 to 2019. In the first approach, we equate the \gls{el} scores using \gls{gke} for nonequivalent groups, as described in Section~\ref{sec2.1:kernel covariate equating}. A moment-based log-linear model is used to pre-smooth the test scores (including polynomial terms of the score up to degree 6, main effects of the covariates, and their interactions with the score), and a Gaussian kernel is applied to continuize the score distributions. The covariates used are the school type (academic or non-academic), attempt to pass the exam (first or repeated), and the score on the \gls{nl} mandatory test (divided into five fixed categories using thresholds of 50, 60, 70, 80, and 100 percent of the maximum score). The data do not include any other background variables. The spring term of 2017 is used as the baseline test form.

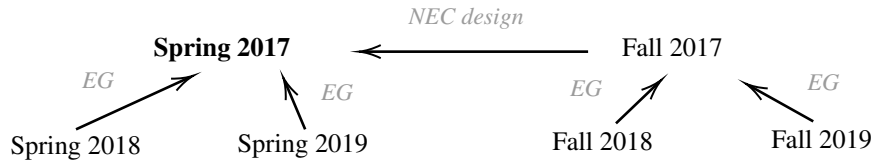
\begin{figure}[h!]
\centering
\tikzset{every picture/.style={line width=1pt}} %set default line width to 0.75pt        

\begin{tikzpicture}[x=0.75pt,y=0.75pt,yscale=-1,xscale=1]
uncomment if require: \path (320,100); %set diagram left start at 0, and has height of 300

%Straight Lines [id:da6352097473405753] 
\draw    (380.6,30.6) -- (267.6,30.6) ;
\draw [shift={(265.6,30.6)}, rotate = 360] [color={rgb, 255:red, 0; green, 0; blue, 0 }  ][line width=0.75]    (10.93,-3.29) .. controls (6.95,-1.4) and (3.31,-0.3) .. (0,0) .. controls (3.31,0.3) and (6.95,1.4) .. (10.93,3.29)   ;
%Straight Lines [id:da0075623955448002045] 
\draw    (238.6,69.6) -- (228.38,45.44) ;
\draw [shift={(227.6,43.6)}, rotate = 67.07] [color={rgb, 255:red, 0; green, 0; blue, 0 }  ][line width=0.75]    (10.93,-3.29) .. controls (6.95,-1.4) and (3.31,-0.3) .. (0,0) .. controls (3.31,0.3) and (6.95,1.4) .. (10.93,3.29)   ;
%Straight Lines [id:da4289380831909203] 
\draw    (123.6,69.6) -- (180.78,43.43) ;
\draw [shift={(182.6,42.6)}, rotate = 155.41] [color={rgb, 255:red, 0; green, 0; blue, 0 }  ][line width=0.75]    (10.93,-3.29) .. controls (6.95,-1.4) and (3.31,-0.3) .. (0,0) .. controls (3.31,0.3) and (6.95,1.4) .. (10.93,3.29)   ;
%Straight Lines [id:da611122343654272] 
\draw    (394.6,66.6) -- (416.15,45.98) ;
\draw [shift={(417.6,44.6)}, rotate = 136.27] [color={rgb, 255:red, 0; green, 0; blue, 0 }  ][line width=0.75]    (10.93,-3.29) .. controls (6.95,-1.4) and (3.31,-0.3) .. (0,0) .. controls (3.31,0.3) and (6.95,1.4) .. (10.93,3.29)   ;
%Straight Lines [id:da3405597878880997] 
\draw    (493.6,65.6) -- (461.35,47.58) ;
\draw [shift={(459.6,46.6)}, rotate = 29.2] [color={rgb, 255:red, 0; green, 0; blue, 0 }  ][line width=0.75]    (10.93,-3.29) .. controls (6.95,-1.4) and (3.31,-0.3) .. (0,0) .. controls (3.31,0.3) and (6.95,1.4) .. (10.93,3.29)   ;

% Text Node
\draw (161,22) node [anchor=north west][inner sep=0.75pt]   [align=left] {\textbf{Spring 2017}};
% Text Node
\draw (396,22) node [anchor=north west][inner sep=0.75pt]   [align=left] {Fall 2017};
% Text Node
\draw (89,71) node [anchor=north west][inner sep=0.75pt]   [align=left] {Spring 2018};
% Text Node
\draw (361,69) node [anchor=north west][inner sep=0.75pt]   [align=left] {Fall 2018};
% Text Node
\draw (203,70) node [anchor=north west][inner sep=0.75pt]   [align=left] {Spring 2019};
% Text Node
\draw (471,68) node [anchor=north west][inner sep=0.75pt]   [align=left] {Fall 2019};
% Text Node
\draw (125,39) node [anchor=north west][inner sep=0.75pt]  [color={rgb, 255:red, 155; green, 155; blue, 155 }  ,opacity=1 ] [align=left] {\textit{{\small EG}}};
% Text Node
\draw (245,45) node [anchor=north west][inner sep=0.75pt]  [color={rgb, 255:red, 155; green, 155; blue, 155 }  ,opacity=1 ] [align=left] {\textit{{\small EG}}};
% Text Node
\draw (370,43) node [anchor=north west][inner sep=0.75pt]  [color={rgb, 255:red, 155; green, 155; blue, 155 }  ,opacity=1 ] [align=left] {\textit{{\small EG}}};
% Text Node
\draw (489,40) node [anchor=north west][inner sep=0.75pt]  [color={rgb, 255:red, 155; green, 155; blue, 155 }  ,opacity=1 ] [align=left] {\textit{{\small EG}}};
% Text Node
\draw (289,7) node [anchor=north west][inner sep=0.75pt]   [align=left] {\textit{{\small \textcolor[rgb]{0.61,0.61,0.61}{NEC design}}}};

\end{tikzpicture}

\caption{The process of equating \gls{nl} test scores across examination terms and years. }
\label{diagram_real_data_example}
\end{figure}

In the second approach, we apply sequential \gls{gke} described in Section \ref{sec:RKCE} to account for the fact that the \gls{nl} test score, used as a covariate, may not be directly comparable across terms, as it is also measured using different test forms. The second approach thus differs from the first approach in that it uses adjusted (instead of raw) \gls{nl} scores with the adjustment reflecting the specific context of individual test administrations. First, using the \gls{eg} design \parencite{kolen2004test, vonDavier2013observed}, we equated \gls{nl} scores from all spring terms to the baseline form spring 2017. The assumption of equivalent groups is plausible in this context because the \gls{nl} test is mandatory for all students and therefore the examinee population is essentially complete each year. Moreover, neither the overall student composition in the country nor the curriculum changed during the period under consideration. 
Consequently, the population of \gls{nl} examinees is expected to be relatively stable across administrations within the spring examination term across years, as also demonstrated by score distributions in Table~\ref{tab:average-test-scores}. Second, using the \gls{eg} design, we equated all fall terms to the baseline form fall 2017. While the populations of fall examinees may not be strictly comparable across years, we consider equating all fall terms using the \gls{eg} design more reasonable than other possible options, given the limited availability of covariates in the empirical datasets. Third, to account for the substantial differences between spring and fall examinees, we applied equating under the \gls{nec} design to equate fall 2017 to spring 2017 using school type and exam attempt as covariates. For clarity, the process of equating \gls{nl} test scores across examination terms and years is illustrated in Figure~\ref{diagram_real_data_example}.  After equating the \gls{nl} test scores, we applied equating under the \gls{nec} design for the \gls{el} test scores, using equated \gls{nl} test score, school type, and exam attempt as covariates. \\

We compare the resulting equated scores for \gls{el} obtained (a) without equating the \gls{nl} test scores and (b) with equating, focusing on two main components: the equated scores and the corresponding \gls{see}s. 

%------------------------------------------------------------

\subsection{Results of real data example}
We applied the equating methodology to real assessment data from six administrations, covering the period from spring 2017 to fall 2019. Although the six test forms are designed to be of similar difficulty, the distribution of total test scores differed, especially in the spring and fall terms (Table~\ref{tab:average-test-scores}), reflecting differences in characteristics of test takers. 

\begin{figure}[h!]
    \centering
    \includegraphics[width=\textwidth]{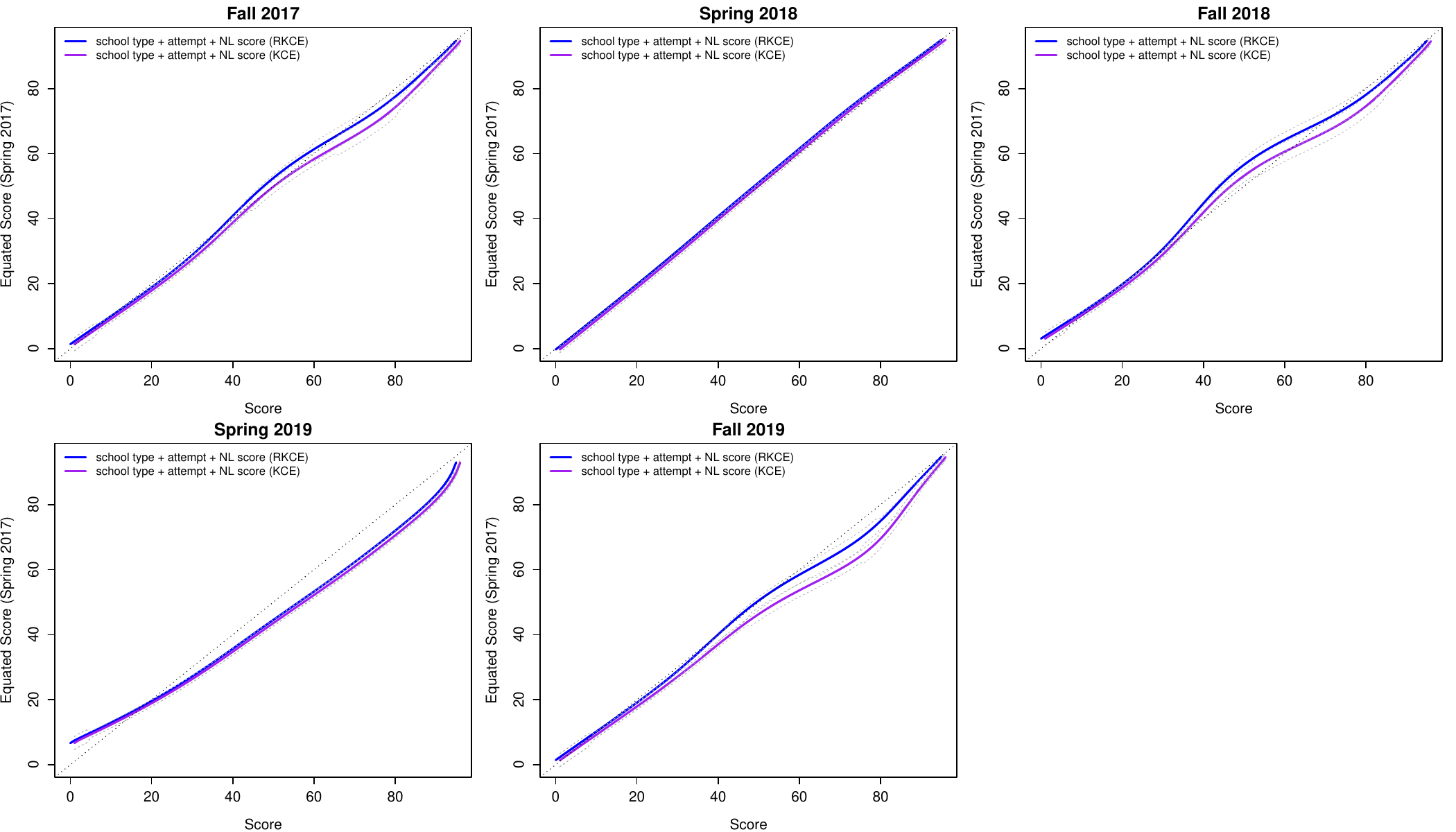}
    \caption{Equated English Language test scores across five testing terms (Fall 2017 – Fall 2019) using models with and without equated \acrlong{nl} (\gls{nl}) test scores. The solid blue lines represent equated scores using school type, exam attempt, and equated NL scores; the dot-dashed purple lines represent equated scores using raw \gls{nl} scores. Dashed gray lines show \gls{see} bounds around each score estimate. The diagonal dotted line indicates the identity line for reference.}
    \label{fig:equated-scores}
\end{figure}

Figure~\ref{fig:equated-scores} displays the equated \gls{el} scores for each term, using spring 2017 as the baseline form. Two equating models were evaluated: one that included school type, exam attempt, and equated \gls{nl} scores, and another that used school type, exam attempt, and raw (non-equated) \gls{nl} scores. Across all terms, the resulting equated \gls{el} scores were very similar, with only minor differences between the two models. This suggests that, in this particular dataset, using equated \gls{nl} scores instead of raw scores has a limited influence on the final equated outcomes. A likely explanation for this minimal impact is the relative stability of the \gls{nl} test forms over time. 

\begin{figure}[h!]
    \centering
    \includegraphics[width=\textwidth]{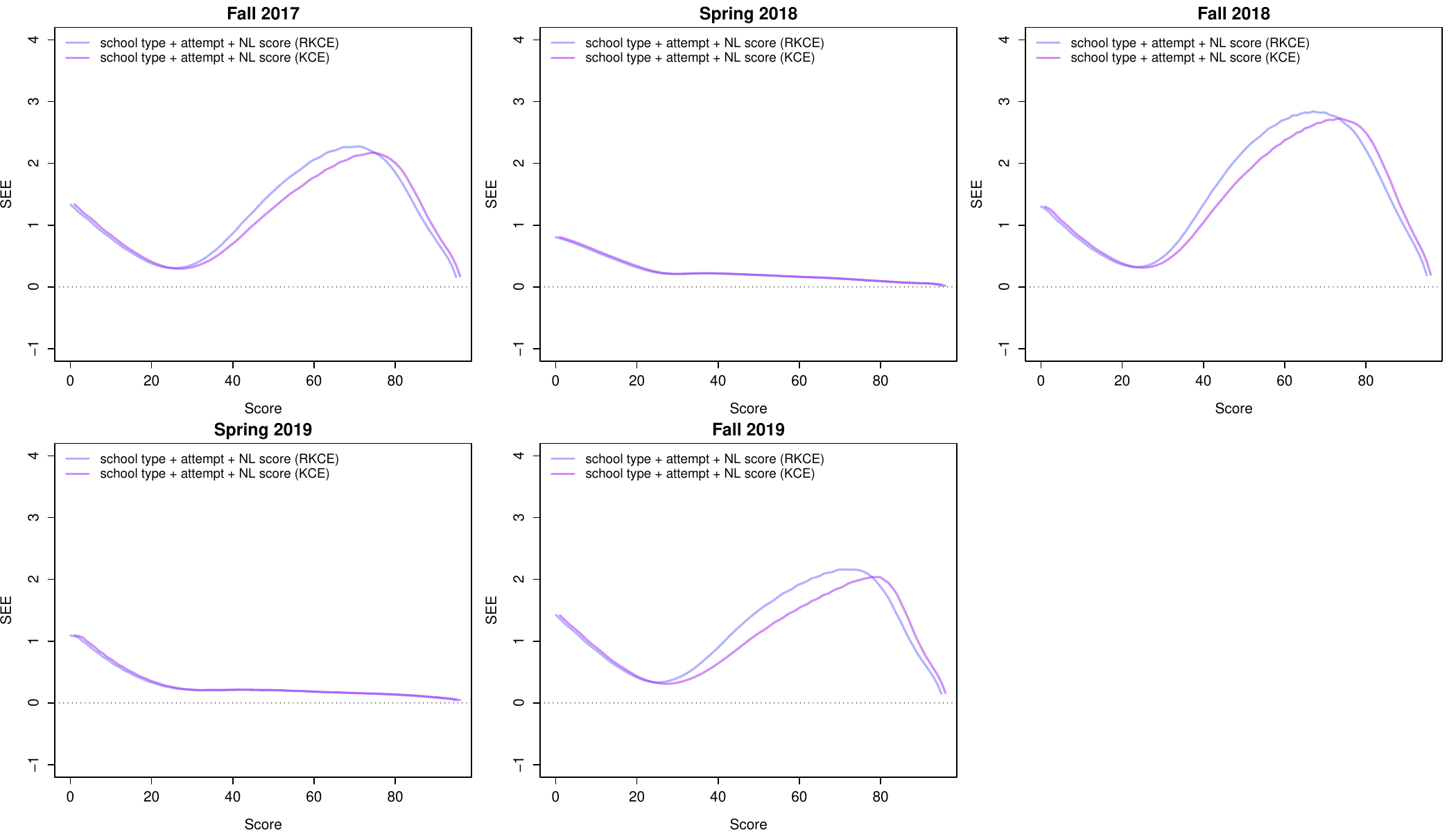}
    \caption{\gls{see} across five testing terms (Fall 2017 – Fall 2019), comparing models using equated versus non-equated \acrlong{nl} (\gls{nl}) test scores. Solid light blue lines represent \gls{see} values for equated scores based on school type, exam attempt, and equated \gls{nl} score; dot-dashed light purple lines represent \gls{see} values when using non-equated \gls{nl} scores. The dotted horizontal line indicates zero for reference.}
    \label{fig:SEE}
\end{figure}

Figure~\ref{fig:SEE} presents the \gls{see} estimates for each model across all terms. 
The model using equated \gls{nl} scores produced slightly higher \gls{see}s in the middle of the score range; this outcome is expected, given that using equated scores as predictors introduces additional variability. Importantly, however, the increase in \gls{see} estimates was minimal compared to \gls{gke}; even with the added variability, the sequential \gls{gke} produced stable and reliable equating results across all test administrations. 

Appendix provides detailed tables for each term, including observed scores, equated scores from both models, and their corresponding \gls{see} values (Tables~\ref{tab:spring_multirow_years} and \ref{tab:falls_multirow_years}).

\begin{figure}[h!]
    \centering
    \includegraphics[width=0.65\textwidth]{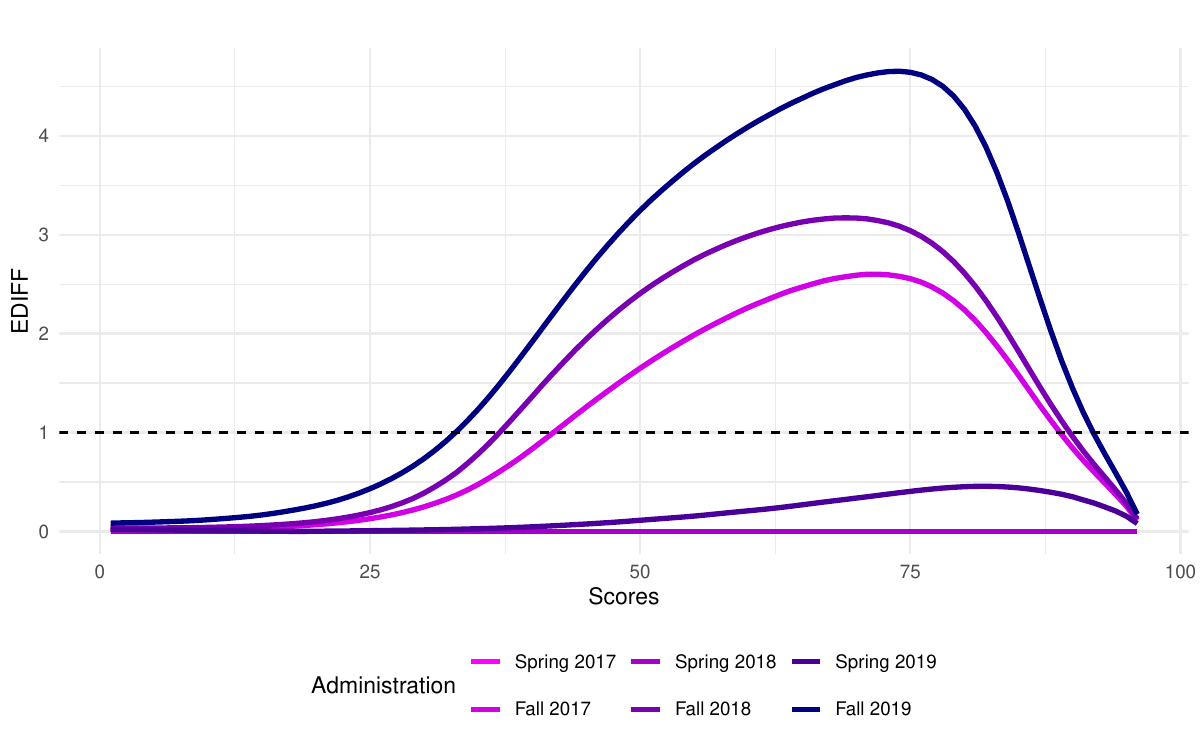}
    \caption{\glsfirst{ediff} between \glsfirst{gke} and Sequential \gls{gke} across the score scale for each administration.}
    \label{fig:EDIFF}
\end{figure}

\begin{figure}[h!]
    \centering
    \includegraphics[width=0.65\textwidth]{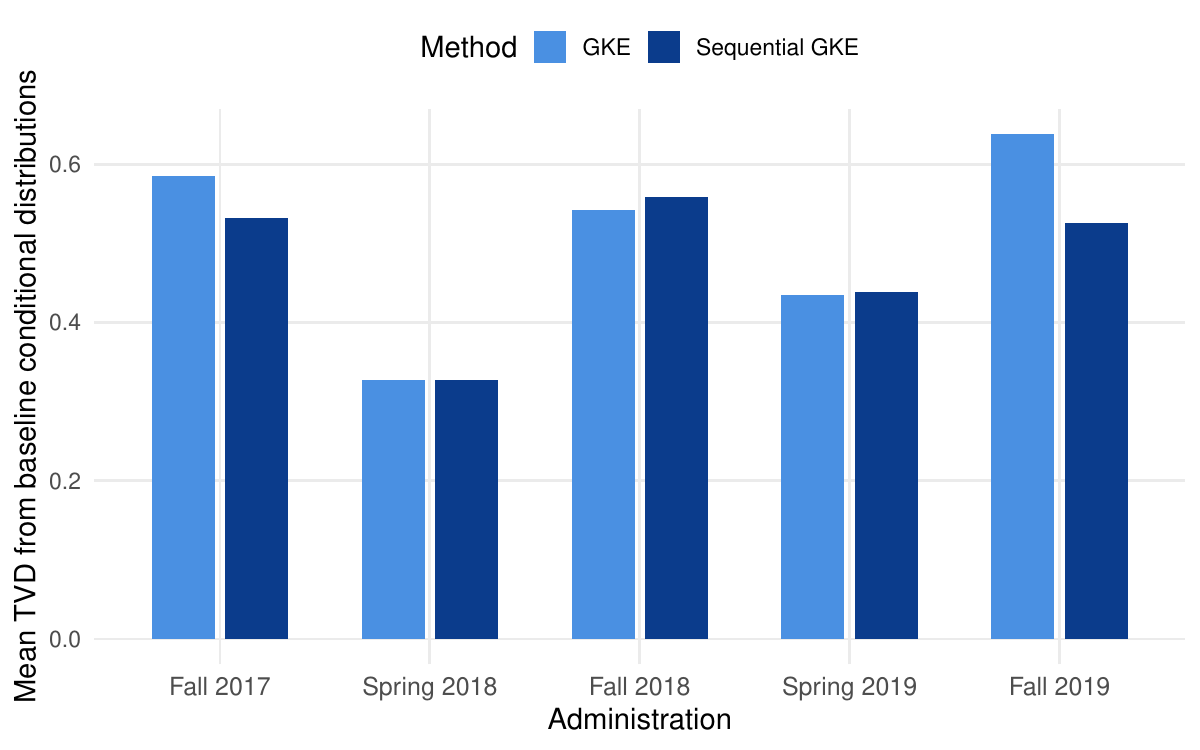}
    \caption{%\glsfirst{tvd}
    Total variation distance (\gls{tvd}) between each administration’s conditional score distributions and those of the baseline administration (Spring 2017). For each administration, \gls{tvd} was computed separately within each covariate group, and the plotted value represents the mean \gls{tvd} across all groups.}
    \label{fig:TV}
\end{figure}

Additionally, to illustrate the differences between the two approaches, Figure~\ref{fig:EDIFF} presents \gls{ediff} (defined in Equation~\eqref{ediff}), which quantifies discrepancies between the \gls{gke} and sequential \gls{gke} equating results across the score scale for each administration. Differences were generally modest, though somewhat larger deviations occurred at the upper end of the scale. To further evaluate whether sequential \gls{gke} enhances invariance and reduces differences in conditional score distributions, the \gls{tvd} was computed to measure the similarity between each administration’s conditional score distributions and those of the baseline administration (Spring 2017). For each administration, \gls{tvd} was calculated within each covariate group and then averaged across groups. As shown in Figure~\ref{fig:TV}, sequential \gls{gke} generally produced slightly lower \gls{tvd} values than \gls{gke} across most administrations, suggesting that incorporating the sequential adjustment can modestly reduce differences in conditional score distributions.

\begin{table}
\caption{Average test scores, standard deviations, and sample sizes by exam term
}\label{tab:average-test-scores}
\begin{center} 
\begin{tabular}{ l r r r r r r }
\toprule
   \multirow{2}{*}{Term} & \multicolumn{3}{c}{\gls{nl}} & \multicolumn{3}{c}{English} \\
   \cmidrule(lr){2-4} \cmidrule(lr){5-7}
        & \multicolumn{1}{c}{Mean} & \multicolumn{1}{c}{SD} & \multicolumn{1}{c}{N} & \multicolumn{1}{c}{Mean} & \multicolumn{1}{c}{SD} & \multicolumn{1}{c}{N} \\
   \midrule
   \noalign{\vskip 1mm}
    Spring 2017 & 32.14 & 8.47 & 64 530 & 71.40 & 17.58 &  44 310\\
    Fall 2017   & 24.06 & 6.56 &  4 880 & 49.14 & 18.25 &   9 176\\
    Spring 2018 & 31.81 & 8.37 & 64 941 & 70.22 & 17.51 &  46 232\\
    Fall 2018   & 24.80 & 5.70 &   5 037 & 47.29 & 18.59 &   9 227\\
    Spring 2019 & 31.64 & 8.37 & 65 772 & 77.42 & 17.49 &  48 798\\
    Fall 2019   & 22.91 & 6.20 &   3 924 & 52.46 & 20.06 &  9 403\\
  \bottomrule
\end{tabular} 
\\
\vskip 0.2em
\footnotesize{Note: Maximum score in the \glsfirst{nl} test is 50; maximum score in the English test is 95.}
\end{center} 
\end{table}

%------------------------------------------------------------
\section{Discussion and conclusion}
%------------------------------------------------------------

This study investigated how violations of the assumption of population-invariant conditional distributions affect the accuracy of equated test scores in the \gls{nec} design. When the covariates used to account for ability differences between the groups are measured with divergent test forms, this assumption may be violated. To reduce the threat of biased equated test scores, we propose sequential \gls{gke} as an extension within the  \gls{gke} framework under the \gls{nec} design \parencite{wiberg2025generalized}, designed to align covariate measurement scales in contexts where no anchor test is administered and where some background characteristics are obtained through distinct measurement tools. The evaluation of this approach, through both a simulation study and an application to real-world data from the national high school leaving exam, offers insights into the practical considerations of applying equating to nonequivalent groups. 

The simulation study clearly demonstrated that when the assumption of population-invariant conditional distributions holds -- specifically, when covariates are measured equivalently across groups -- the bias in equated scores using the standard \gls{gke} approach is minimal and tends to decrease with increasing sample size. However, when this assumption is violated -- particularly due to covariates measured on misaligned scales used to measure a continuous covariate -- the resulting bias increases substantially, especially when the covariate is strongly related to the total test scores being equated. This finding highlights the sensitivity of equating with the \gls{nec} design to violations of measurement invariance in the covariates themselves. In such cases, sequential \gls{gke} substantially reduces bias compared to \gls{gke}.

The proposed sequential \gls{gke} approach was illustrated with a real data example involving English and Native Language (\gls{nl}) tests administered over multiple years and terms. Sequential \gls{gke} with equated \gls{nl} scores led to minor changes in equated \gls{el} scores compared to standard \gls{gke} using raw \gls{nl} scores as a covariate. This suggests that the \gls{nl} tests were largely stable over time, showing consistent difficulty across administrations. As a result, the extent of measurement invariance violations that motivated the development of sequential \gls{gke} appeared limited in this dataset.
The observed increase in \gls{see}s when using equated \gls{nl} scores was expected, reflecting the additional variability introduced by the step of equating the covariate. Despite this, the overall \gls{see} increase was minimal, and the equating results remained stable and reliable, supporting the practical utility of sequential \gls{gke} even in settings where violations of covariate invariance are relatively minor.

Our findings have several practical implications for test equating in high-stakes educational assessments. First, when covariates are known or suspected to be measured on different scales or with varying difficulty across groups, sequential \gls{gke} offers a methodological solution to mitigate bias. This sequential equating approach allows practitioners to adjust for differences in covariate measurement prior to the main equating step, thereby enhancing fairness and score comparability.
Second, in settings where covariates are stable and equivalent across groups -- such as the \gls{nl} test in the national high school leaving exam -- using raw covariates may be sufficient, simplifying the equating process without compromising accuracy. However, practitioners should empirically verify this assumption, since failing to account for covariate measurement differences can introduce substantial bias, as demonstrated in the simulation study.

While we demonstrated sequential \gls{gke} using an example from a national high school leaving exam, the proposed equating approach has broader potential applications. For example, employing disparate scales to measure the same underlying construct in different populations is common in psychology and mental health research \parencite{jovic2024psychology}. More generally, researchers in psychology, health outcomes research, and the social sciences frequently seek to establish a common interpretive scale across instruments measuring the same or closely related latent constructs. In such settings, the objective is often not strict test equating but rather score linking, cross-walking, or measurement harmonization \parencite{dorans2004equating,deBeurs2022harmonize,tang2022crosswalk}. In such cases, scale equating is typically performed within a single-group design based on a calibration sample \parencite{huang2022calibration, jongsma2022calibration}. When calibration data are unavailable, equating with the \gls{nec} design may serve as a cost-effective alternative -- particularly when researchers have collected additional participant information. If some covariates are themselves measured with different test forms, or with non-equivalent instruments intended to assess the same construct, sequential \gls{gke} provides a suitable approach for addressing discrepancies in covariate measurement in psychology and related fields. Furthermore, the assumption of population-invariant conditional distributions, while central to the \gls{nec} design, also plays an important role in related equating frameworks, such as the \gls{neat} design \parencite{kolen2004test,vonDavier2004kernel}.

Although the present study clearly demonstrated that the proposed sequential \gls{gke} outperforms the standard \gls{gke} approach when the assumption of population-invariant conditional distributions is violated, it is not without limitations. While the simulation design was motivated by realistic parameters, it focused on relatively simple covariate structures (two binary and one continuous variable) and assumed linear relationships. Real-world testing scenarios may involve more complex covariate interactions, nonlinear associations, and multidimensional test constructs, which warrant further investigations.

Additionally, the equating methods were applied here primarily to large-scale testing scenarios. The effects of sample size, particularly in smaller or highly skewed populations, remain an important area for further study. Future research should also explore alternative equating models that better account for complex measurement invariance violations, as well as approaches that integrate covariates with anchor-item methods \parencite{albano2019linking, laukaityte2025combining, wiberg2015kernel}.
Finally, longitudinal applications of sequential \gls{gke} should be investigated to evaluate score comparability over extended time frames and to support the monitoring of educational trends and developmental or behavioral trajectories.

In summary, this study confirms that \gls{gke} is sensitive to violations of the population-invariant conditional distributions assumption, but that sequential \gls{gke} offers an effective methodological extension when differences in covariate measurement are present. The real data application highlights the importance of empirically assessing covariate stability and illustrates how sequential \gls{gke} can maintain equating accuracy even when the covariates themselves must be equated.

%--------------------------------------------------
% SUPPLEMENTARY MATERIAL
%--------------------------------------------------
\section*{Online Supplementary Material}
Supplementary material, including the accompanying \texttt{R} scripts, is available at \href{https://osf.io/kwz3d/}{https://osf.io/kwz3d/}.

%--------------------------------------------------
% Acknowledgements
%--------------------------------------------------

\section*{Acknowledgments and declarations}
The authors thank the Centre for Evaluation of Educational Achievement for providing an anonymized dataset of national school leaving exams containing individual student scores in both the NL and EL tests. We also thank the editors and three anonymous reviewers for their constructive comments on earlier versions of the manuscript. During the preparation of this work, the authors used ChatGPT-4 (OpenAI, 2026) and Claude (Anthropic, 2026) for English proofreading and for language refinement. The authors reviewed and edited the generated content as needed and take full responsibility for the content of the publication.
\section*{Funding}
The study was funded by the Czech Science Foundation project ``Complex analysis of educational measurement data to understand cognitive demands of assessment tasks'' grant number 25-16951S, by the project ``Research of Excellence on Digital Technologies and Wellbeing CZ.02.01.01/00/22\_008/0004583'' which is co-financed by the European Union, by the institutional support RVO 67985807, and by the Charles University, project GA UK No. 117224. 

%--------------------------------------------------
% BIBLIOGRAPHY
%--------------------------------------------------

%------------------------------------------------------------
\section*{References}
%------------------------------------------------------------

\printbibliography[heading = none]

\newpage

%------------------------------------------------------------
\section*{Appendix}
%------------------------------------------------------------

\captionsetup{font=tiny} 
\begin{table}[ht]
\vspace{-2cm}
\centering
\begin{tiny}
\begin{tabular}{c
                cc
                cc c c c c}
\toprule
\multirow{7}{*}{Score} & \multicolumn{4}{c}{Spring 2018} & \multicolumn{4}{c}{Spring 2019} \\
\cmidrule(lr){2-5} \cmidrule(lr){6-9}
                        & \multicolumn{2}{c}{GKE} & \multicolumn{2}{c}{Sequential GKE} &  \multicolumn{2}{c}{GKE} & \multicolumn{2}{c}{Sequential GKE} \\
                        & Eq. scores & SEE     & Eq. scores & SEE 
                        & Eq. scores & SEE 
                        & Eq. scores & SEE \\
\midrule
0 & -0.25 & 0.81 & -0.25 & 0.81 & 6.68 & 1.09 & 6.66 & 1.09 \\ 
    1 & 0.74 & 0.79 & 0.74 & 0.79 & 7.47 & 1.08 & 7.45 & 1.08 \\ 
    2 & 1.73 & 0.77 & 1.73 & 0.77 & 8.11 & 1.06 & 8.09 & 1.06 \\ 
    3 & 2.73 & 0.75 & 2.73 & 0.75 & 8.70 & 1.00 & 8.69 & 1.00 \\ 
    4 & 3.72 & 0.72 & 3.72 & 0.72 & 9.30 & 0.95 & 9.28 & 0.95 \\ 
    5 & 4.72 & 0.70 & 4.72 & 0.70 & 9.89 & 0.90 & 9.88 & 0.90 \\ 
    6 & 5.71 & 0.67 & 5.71 & 0.67 & 10.50 & 0.84 & 10.48 & 0.84 \\ 
    7 & 6.71 & 0.64 & 6.71 & 0.64 & 11.10 & 0.80 & 11.09 & 0.80 \\ 
    8 & 7.71 & 0.62 & 7.71 & 0.62 & 11.71 & 0.74 & 11.70 & 0.74 \\ 
    9 & 8.71 & 0.59 & 8.71 & 0.59 & 12.33 & 0.71 & 12.32 & 0.71 \\ 
   10 & 9.71 & 0.56 & 9.71 & 0.56 & 12.95 & 0.66 & 12.94 & 0.66 \\ 
   11 & 10.71 & 0.54 & 10.71 & 0.54 & 13.59 & 0.62 & 13.58 & 0.62 \\ 
   12 & 11.71 & 0.51 & 11.71 & 0.51 & 14.22 & 0.58 & 14.22 & 0.58 \\ 
   13 & 12.72 & 0.49 & 12.72 & 0.49 & 14.87 & 0.54 & 14.86 & 0.54 \\ 
   14 & 13.72 & 0.46 & 13.72 & 0.46 & 15.52 & 0.50 & 15.52 & 0.50 \\ 
   15 & 14.73 & 0.44 & 14.73 & 0.44 & 16.18 & 0.47 & 16.18 & 0.47 \\ 
   16 & 15.73 & 0.41 & 15.73 & 0.41 & 16.85 & 0.44 & 16.85 & 0.44 \\ 
   17 & 16.74 & 0.39 & 16.74 & 0.39 & 17.53 & 0.41 & 17.53 & 0.41 \\ 
   18 & 17.75 & 0.36 & 17.75 & 0.36 & 18.21 & 0.38 & 18.21 & 0.38 \\ 
   19 & 18.77 & 0.34 & 18.77 & 0.34 & 18.90 & 0.35 & 18.90 & 0.35 \\ 
   20 & 19.78 & 0.32 & 19.78 & 0.32 & 19.60 & 0.33 & 19.60 & 0.33 \\ 
   21 & 20.80 & 0.29 & 20.80 & 0.29 & 20.31 & 0.31 & 20.31 & 0.31 \\ 
   22 & 21.82 & 0.28 & 21.82 & 0.28 & 21.03 & 0.29 & 21.03 & 0.29 \\ 
   23 & 22.85 & 0.26 & 22.85 & 0.26 & 21.76 & 0.27 & 21.76 & 0.27 \\ 
   24 & 23.87 & 0.24 & 23.87 & 0.24 & 22.50 & 0.26 & 22.50 & 0.26 \\ 
   25 & 24.91 & 0.23 & 24.91 & 0.23 & 23.24 & 0.25 & 23.25 & 0.25 \\ 
   26 & 25.94 & 0.22 & 25.94 & 0.22 & 24.00 & 0.23 & 24.01 & 0.23 \\ 
   27 & 26.98 & 0.22 & 26.98 & 0.22 & 24.77 & 0.23 & 24.78 & 0.23 \\ 
   28 & 28.02 & 0.21 & 28.02 & 0.21 & 25.55 & 0.22 & 25.56 & 0.22 \\ 
   29 & 29.07 & 0.21 & 29.07 & 0.21 & 26.34 & 0.22 & 26.35 & 0.22 \\ 
   30 & 30.12 & 0.21 & 30.12 & 0.21 & 27.14 & 0.21 & 27.15 & 0.21 \\ 
   31 & 31.17 & 0.21 & 31.17 & 0.21 & 27.95 & 0.21 & 27.97 & 0.21 \\ 
   32 & 32.22 & 0.21 & 32.22 & 0.21 & 28.77 & 0.21 & 28.79 & 0.21 \\ 
   33 & 33.28 & 0.22 & 33.28 & 0.22 & 29.60 & 0.21 & 29.62 & 0.21 \\ 
   34 & 34.34 & 0.22 & 34.34 & 0.22 & 30.44 & 0.21 & 30.47 & 0.21 \\ 
   35 & 35.40 & 0.22 & 35.40 & 0.22 & 31.29 & 0.21 & 31.32 & 0.21 \\ 
   36 & 36.46 & 0.22 & 36.46 & 0.22 & 32.14 & 0.21 & 32.18 & 0.21 \\ 
   37 & 37.52 & 0.22 & 37.52 & 0.22 & 33.01 & 0.21 & 33.05 & 0.21 \\ 
   38 & 38.58 & 0.22 & 38.58 & 0.22 & 33.88 & 0.21 & 33.92 & 0.21 \\ 
   39 & 39.63 & 0.22 & 39.63 & 0.22 & 34.76 & 0.21 & 34.80 & 0.21 \\ 
   40 & 40.69 & 0.22 & 40.69 & 0.22 & 35.64 & 0.21 & 35.69 & 0.21 \\ 
   41 & 41.74 & 0.22 & 41.74 & 0.22 & 36.52 & 0.21 & 36.58 & 0.21 \\ 
   42 & 42.80 & 0.21 & 42.80 & 0.21 & 37.40 & 0.21 & 37.47 & 0.21 \\ 
   43 & 43.84 & 0.21 & 43.84 & 0.21 & 38.29 & 0.21 & 38.36 & 0.21 \\ 
   44 & 44.89 & 0.21 & 44.89 & 0.21 & 39.17 & 0.21 & 39.25 & 0.21 \\ 
   45 & 45.94 & 0.21 & 45.94 & 0.21 & 40.06 & 0.21 & 40.14 & 0.21 \\ 
   46 & 46.98 & 0.20 & 46.98 & 0.20 & 40.94 & 0.21 & 41.03 & 0.21 \\ 
   47 & 48.02 & 0.20 & 48.02 & 0.20 & 41.82 & 0.21 & 41.92 & 0.21 \\ 
   48 & 49.06 & 0.20 & 49.06 & 0.20 & 42.70 & 0.21 & 42.80 & 0.21 \\ 
   49 & 50.10 & 0.20 & 50.10 & 0.20 & 43.57 & 0.21 & 43.69 & 0.21 \\ 
   50 & 51.14 & 0.19 & 51.14 & 0.19 & 44.44 & 0.21 & 44.56 & 0.21 \\ 
   51 & 52.18 & 0.19 & 52.18 & 0.19 & 45.31 & 0.20 & 45.44 & 0.21 \\ 
   52 & 53.22 & 0.19 & 53.22 & 0.19 & 46.17 & 0.20 & 46.31 & 0.20 \\ 
   53 & 54.25 & 0.18 & 54.25 & 0.18 & 47.04 & 0.20 & 47.18 & 0.20 \\ 
   54 & 55.29 & 0.18 & 55.29 & 0.18 & 47.90 & 0.20 & 48.05 & 0.20 \\ 
   55 & 56.33 & 0.18 & 56.33 & 0.18 & 48.76 & 0.19 & 48.92 & 0.19 \\ 
   56 & 57.37 & 0.17 & 57.37 & 0.17 & 49.62 & 0.19 & 49.79 & 0.19 \\ 
   57 & 58.41 & 0.17 & 58.41 & 0.17 & 50.47 & 0.19 & 50.66 & 0.19 \\ 
   58 & 59.46 & 0.17 & 59.46 & 0.17 & 51.33 & 0.19 & 51.53 & 0.19 \\ 
   59 & 60.50 & 0.17 & 60.50 & 0.17 & 52.19 & 0.18 & 52.40 & 0.19 \\ 
   60 & 61.54 & 0.16 & 61.54 & 0.16 & 53.05 & 0.18 & 53.27 & 0.18 \\ 
   61 & 62.58 & 0.16 & 62.58 & 0.16 & 53.92 & 0.18 & 54.15 & 0.18 \\ 
   62 & 63.62 & 0.16 & 63.62 & 0.16 & 54.79 & 0.18 & 55.03 & 0.18 \\ 
   63 & 64.67 & 0.16 & 64.67 & 0.16 & 55.66 & 0.17 & 55.92 & 0.17 \\ 
   64 & 65.70 & 0.15 & 65.70 & 0.15 & 56.54 & 0.17 & 56.81 & 0.17 \\ 
   65 & 66.74 & 0.15 & 66.74 & 0.15 & 57.42 & 0.17 & 57.71 & 0.17 \\ 
   66 & 67.77 & 0.15 & 67.77 & 0.15 & 58.31 & 0.17 & 58.61 & 0.17 \\ 
   67 & 68.80 & 0.14 & 68.80 & 0.14 & 59.21 & 0.16 & 59.52 & 0.17 \\ 
   68 & 69.82 & 0.14 & 69.82 & 0.14 & 60.11 & 0.16 & 60.44 & 0.17 \\ 
   69 & 70.83 & 0.14 & 70.83 & 0.14 & 61.02 & 0.16 & 61.36 & 0.16 \\ 
   70 & 71.84 & 0.13 & 71.84 & 0.13 & 61.94 & 0.16 & 62.29 & 0.16 \\ 
   71 & 72.83 & 0.13 & 72.83 & 0.13 & 62.87 & 0.16 & 63.23 & 0.16 \\ 
   72 & 73.82 & 0.12 & 73.82 & 0.12 & 63.80 & 0.15 & 64.18 & 0.16 \\ 
   73 & 74.80 & 0.12 & 74.80 & 0.12 & 64.74 & 0.15 & 65.13 & 0.15 \\ 
   74 & 75.77 & 0.12 & 75.77 & 0.12 & 65.69 & 0.15 & 66.09 & 0.15 \\ 
   75 & 76.73 & 0.11 & 76.73 & 0.11 & 66.64 & 0.15 & 67.06 & 0.15 \\ 
   76 & 77.67 & 0.11 & 77.67 & 0.11 & 67.61 & 0.15 & 68.04 & 0.15 \\ 
   77 & 78.61 & 0.10 & 78.61 & 0.10 & 68.58 & 0.14 & 69.02 & 0.14 \\ 
   78 & 79.54 & 0.10 & 79.54 & 0.10 & 69.56 & 0.14 & 70.00 & 0.14 \\ 
   79 & 80.46 & 0.10 & 80.46 & 0.10 & 70.54 & 0.14 & 71.00 & 0.14 \\ 
   80 & 81.38 & 0.09 & 81.38 & 0.09 & 71.54 & 0.14 & 72.00 & 0.13 \\ 
   81 & 82.29 & 0.09 & 82.29 & 0.09 & 72.55 & 0.13 & 73.00 & 0.13 \\ 
   82 & 83.19 & 0.08 & 83.19 & 0.08 & 73.57 & 0.13 & 74.02 & 0.13 \\ 
   83 & 84.09 & 0.08 & 84.09 & 0.08 & 74.60 & 0.12 & 75.05 & 0.12 \\ 
   84 & 84.99 & 0.08 & 84.99 & 0.08 & 75.65 & 0.12 & 76.09 & 0.12 \\ 
   85 & 85.89 & 0.07 & 85.89 & 0.07 & 76.72 & 0.11 & 77.15 & 0.11 \\ 
   86 & 86.78 & 0.07 & 86.78 & 0.07 & 77.82 & 0.11 & 78.24 & 0.11 \\ 
   87 & 87.68 & 0.07 & 87.68 & 0.07 & 78.96 & 0.10 & 79.35 & 0.10 \\ 
   88 & 88.58 & 0.06 & 88.58 & 0.06 & 80.14 & 0.10 & 80.51 & 0.10 \\ 
   89 & 89.48 & 0.06 & 89.48 & 0.06 & 81.38 & 0.09 & 81.73 & 0.09 \\ 
   90 & 90.39 & 0.06 & 90.39 & 0.06 & 82.70 & 0.09 & 83.02 & 0.09 \\ 
   91 & 91.30 & 0.06 & 91.30 & 0.06 & 84.13 & 0.08 & 84.41 & 0.08 \\ 
   92 & 92.22 & 0.05 & 92.22 & 0.05 & 85.71 & 0.08 & 85.96 & 0.07 \\ 
   93 & 93.15 & 0.05 & 93.15 & 0.05 & 87.53 & 0.07 & 87.74 & 0.07 \\ 
   94 & 94.08 & 0.04 & 94.08 & 0.04 & 89.77 & 0.06 & 89.92 & 0.06 \\ 
   95 & 95.03 & 0.02 & 95.03 & 0.02 & 92.91 & 0.04 & 92.99 & 0.04 \\ 
\bottomrule
\end{tabular}
\end{tiny}
\vspace{-0.5em}
\caption{\tiny Equated scores and \gls{see} for spring sessions using \gls{gke} and Sequential \gls{gke} across years.}
\label{tab:spring_multirow_years}
\end{table}

\captionsetup{font=tiny} 
\begin{table}[ht]
\vspace{-2cm}
\centering
\begin{tiny}
\begin{tabular}{c
                cc
                cc c c c c c c c c}
\toprule
\multirow{7}{*}{Score} & \multicolumn{4}{c}{Fall 2017} & \multicolumn{4}{c}{Fall 2018} & \multicolumn{4}{c}{Fall 2019} \\
\cmidrule(lr){2-5} \cmidrule(lr){6-9}  \cmidrule(lr){10-13}
                        & \multicolumn{2}{c}{GKE} & \multicolumn{2}{c}{Sequential GKE} &  \multicolumn{2}{c}{GKE} & \multicolumn{2}{c}{Sequential GKE} &  \multicolumn{2}{c}{GKE} & \multicolumn{2}{c}{Sequential GKE} \\
                        & Eq. scores & SEE     & Eq. scores & SEE 
                        & Eq. scores & SEE 
                        & Eq. scores & SEE & Eq. scores & SEE 
                        & Eq. scores & SEE \\
\midrule
0 & 1.39 & 1.34 & 1.40 & 1.34 & 3.07 & 1.30 & 3.10 & 1.30 & 1.34 & 1.42 & 1.42 & 1.42 \\ 
    1 & 2.26 & 1.28 & 2.27 & 1.28 & 3.90 & 1.27 & 3.93 & 1.27 & 2.22 & 1.36 & 2.31 & 1.37 \\ 
    2 & 3.12 & 1.22 & 3.13 & 1.21 & 4.69 & 1.21 & 4.72 & 1.21 & 3.10 & 1.29 & 3.18 & 1.30 \\ 
    3 & 3.98 & 1.16 & 4.00 & 1.16 & 5.47 & 1.14 & 5.51 & 1.14 & 3.97 & 1.24 & 4.06 & 1.24 \\ 
    4 & 4.84 & 1.11 & 4.86 & 1.11 & 6.26 & 1.07 & 6.30 & 1.07 & 4.85 & 1.19 & 4.94 & 1.19 \\ 
    5 & 5.70 & 1.06 & 5.72 & 1.06 & 7.06 & 1.02 & 7.09 & 1.02 & 5.72 & 1.13 & 5.82 & 1.13 \\ 
    6 & 6.56 & 0.99 & 6.58 & 1.00 & 7.85 & 0.96 & 7.89 & 0.97 & 6.59 & 1.06 & 6.69 & 1.07 \\ 
    7 & 7.42 & 0.94 & 7.44 & 0.94 & 8.65 & 0.90 & 8.69 & 0.90 & 7.46 & 1.00 & 7.56 & 1.01 \\ 
    8 & 8.28 & 0.89 & 8.30 & 0.89 & 9.45 & 0.84 & 9.49 & 0.84 & 8.33 & 0.95 & 8.44 & 0.95 \\ 
    9 & 9.14 & 0.84 & 9.16 & 0.84 & 10.26 & 0.79 & 10.30 & 0.79 & 9.20 & 0.90 & 9.32 & 0.90 \\ 
   10 & 10.00 & 0.79 & 10.02 & 0.79 & 11.07 & 0.74 & 11.12 & 0.75 & 10.07 & 0.84 & 10.19 & 0.85 \\ 
   11 & 10.86 & 0.74 & 10.89 & 0.74 & 11.89 & 0.69 & 11.94 & 0.69 & 10.93 & 0.79 & 11.07 & 0.80 \\ 
   12 & 11.73 & 0.68 & 11.76 & 0.69 & 12.72 & 0.64 & 12.77 & 0.64 & 11.80 & 0.73 & 11.94 & 0.74 \\ 
   13 & 12.59 & 0.64 & 12.63 & 0.64 & 13.55 & 0.60 & 13.61 & 0.60 & 12.67 & 0.68 & 12.82 & 0.69 \\ 
   14 & 13.46 & 0.60 & 13.50 & 0.60 & 14.40 & 0.56 & 14.45 & 0.56 & 13.54 & 0.64 & 13.71 & 0.64 \\ 
   15 & 14.34 & 0.56 & 14.38 & 0.56 & 15.25 & 0.52 & 15.31 & 0.52 & 14.41 & 0.59 & 14.59 & 0.60 \\ 
   16 & 15.21 & 0.52 & 15.26 & 0.52 & 16.11 & 0.48 & 16.18 & 0.48 & 15.28 & 0.55 & 15.48 & 0.56 \\ 
   17 & 16.10 & 0.48 & 16.15 & 0.48 & 16.98 & 0.44 & 17.06 & 0.44 & 16.15 & 0.51 & 16.37 & 0.52 \\ 
   18 & 16.98 & 0.44 & 17.04 & 0.44 & 17.87 & 0.41 & 17.96 & 0.41 & 17.03 & 0.47 & 17.27 & 0.48 \\ 
   19 & 17.88 & 0.41 & 17.95 & 0.41 & 18.78 & 0.38 & 18.87 & 0.38 & 17.91 & 0.43 & 18.17 & 0.44 \\ 
   20 & 18.78 & 0.38 & 18.86 & 0.38 & 19.69 & 0.36 & 19.81 & 0.36 & 18.80 & 0.40 & 19.08 & 0.41 \\ 
   21 & 19.70 & 0.35 & 19.78 & 0.36 & 20.63 & 0.34 & 20.76 & 0.34 & 19.69 & 0.38 & 20.00 & 0.38 \\ 
   22 & 20.62 & 0.33 & 20.72 & 0.34 & 21.59 & 0.32 & 21.74 & 0.33 & 20.58 & 0.35 & 20.93 & 0.36 \\ 
   23 & 21.55 & 0.32 & 21.67 & 0.32 & 22.57 & 0.32 & 22.74 & 0.32 & 21.48 & 0.34 & 21.87 & 0.35 \\ 
   24 & 22.50 & 0.31 & 22.63 & 0.31 & 23.57 & 0.31 & 23.76 & 0.32 & 22.39 & 0.32 & 22.82 & 0.34 \\ 
   25 & 23.46 & 0.30 & 23.61 & 0.30 & 24.60 & 0.31 & 24.82 & 0.33 & 23.30 & 0.31 & 23.78 & 0.33 \\ 
   26 & 24.44 & 0.30 & 24.60 & 0.30 & 25.66 & 0.32 & 25.91 & 0.34 & 24.23 & 0.31 & 24.76 & 0.34 \\ 
   27 & 25.43 & 0.30 & 25.62 & 0.31 & 26.75 & 0.34 & 27.04 & 0.36 & 25.16 & 0.31 & 25.75 & 0.34 \\ 
   28 & 26.44 & 0.31 & 26.66 & 0.32 & 27.87 & 0.36 & 28.21 & 0.40 & 26.10 & 0.32 & 26.76 & 0.36 \\ 
   29 & 27.47 & 0.32 & 27.72 & 0.34 & 29.02 & 0.39 & 29.41 & 0.44 & 27.05 & 0.33 & 27.79 & 0.38 \\ 
   30 & 28.52 & 0.34 & 28.80 & 0.36 & 30.21 & 0.43 & 30.66 & 0.48 & 28.01 & 0.34 & 28.83 & 0.40 \\ 
   31 & 29.59 & 0.36 & 29.91 & 0.39 & 31.43 & 0.48 & 31.95 & 0.54 & 28.98 & 0.36 & 29.89 & 0.43 \\ 
   32 & 30.68 & 0.39 & 31.05 & 0.42 & 32.68 & 0.53 & 33.28 & 0.60 & 29.96 & 0.39 & 30.97 & 0.47 \\ 
   33 & 31.79 & 0.42 & 32.21 & 0.46 & 33.96 & 0.59 & 34.65 & 0.68 & 30.95 & 0.41 & 32.07 & 0.51 \\ 
   34 & 32.92 & 0.46 & 33.40 & 0.51 & 35.27 & 0.65 & 36.05 & 0.75 & 31.95 & 0.44 & 33.18 & 0.56 \\ 
   35 & 34.07 & 0.50 & 34.61 & 0.56 & 36.60 & 0.73 & 37.48 & 0.84 & 32.95 & 0.48 & 34.31 & 0.61 \\ 
   36 & 35.23 & 0.54 & 35.84 & 0.61 & 37.94 & 0.80 & 38.94 & 0.93 & 33.96 & 0.51 & 35.45 & 0.66 \\ 
   37 & 36.40 & 0.60 & 37.08 & 0.67 & 39.28 & 0.88 & 40.40 & 1.04 & 34.97 & 0.55 & 36.60 & 0.72 \\ 
   38 & 37.59 & 0.65 & 38.34 & 0.73 & 40.62 & 0.96 & 41.86 & 1.13 & 35.98 & 0.60 & 37.75 & 0.78 \\ 
   39 & 38.77 & 0.70 & 39.60 & 0.80 & 41.95 & 1.04 & 43.31 & 1.24 & 36.99 & 0.64 & 38.91 & 0.83 \\ 
   40 & 39.95 & 0.75 & 40.87 & 0.87 & 43.25 & 1.13 & 44.74 & 1.34 & 37.99 & 0.69 & 40.05 & 0.90 \\ 
   41 & 41.13 & 0.81 & 42.13 & 0.94 & 44.53 & 1.22 & 46.13 & 1.43 & 38.98 & 0.73 & 41.19 & 0.96 \\ 
   42 & 42.29 & 0.88 & 43.38 & 1.01 & 45.77 & 1.30 & 47.49 & 1.54 & 39.97 & 0.78 & 42.32 & 1.03 \\ 
   43 & 43.44 & 0.94 & 44.61 & 1.08 & 46.97 & 1.37 & 48.80 & 1.63 & 40.93 & 0.83 & 43.43 & 1.10 \\ 
   44 & 44.57 & 1.00 & 45.83 & 1.15 & 48.12 & 1.45 & 50.06 & 1.72 & 41.88 & 0.88 & 44.52 & 1.16 \\ 
   45 & 45.68 & 1.06 & 47.01 & 1.22 & 49.23 & 1.53 & 51.28 & 1.81 & 42.82 & 0.93 & 45.58 & 1.22 \\ 
   46 & 46.76 & 1.11 & 48.17 & 1.29 & 50.30 & 1.61 & 52.44 & 1.90 & 43.73 & 0.99 & 46.62 & 1.28 \\ 
   47 & 47.81 & 1.17 & 49.31 & 1.36 & 51.31 & 1.68 & 53.55 & 1.99 & 44.62 & 1.04 & 47.63 & 1.34 \\ 
   48 & 48.83 & 1.23 & 50.41 & 1.42 & 52.29 & 1.75 & 54.61 & 2.06 & 45.48 & 1.09 & 48.62 & 1.39 \\ 
   49 & 49.82 & 1.28 & 51.47 & 1.49 & 53.21 & 1.82 & 55.62 & 2.14 & 46.33 & 1.13 & 49.57 & 1.45 \\ 
   50 & 50.79 & 1.34 & 52.51 & 1.55 & 54.10 & 1.88 & 56.58 & 2.21 & 47.15 & 1.17 & 50.50 & 1.50 \\ 
   51 & 51.72 & 1.39 & 53.52 & 1.61 & 54.94 & 1.94 & 57.50 & 2.28 & 47.94 & 1.21 & 51.39 & 1.55 \\ 
   52 & 52.63 & 1.45 & 54.49 & 1.67 & 55.75 & 2.01 & 58.37 & 2.34 & 48.72 & 1.26 & 52.26 & 1.59 \\ 
   53 & 53.51 & 1.50 & 55.43 & 1.73 & 56.52 & 2.07 & 59.21 & 2.39 & 49.47 & 1.31 & 53.11 & 1.63 \\ 
   54 & 54.36 & 1.55 & 56.35 & 1.78 & 57.27 & 2.12 & 60.01 & 2.44 & 50.21 & 1.34 & 53.93 & 1.68 \\ 
   55 & 55.19 & 1.59 & 57.24 & 1.83 & 57.98 & 2.17 & 60.78 & 2.50 & 50.92 & 1.38 & 54.72 & 1.73 \\ 
   56 & 56.00 & 1.63 & 58.10 & 1.87 & 58.67 & 2.23 & 61.52 & 2.56 & 51.62 & 1.43 & 55.50 & 1.77 \\ 
   57 & 56.78 & 1.68 & 58.94 & 1.92 & 59.33 & 2.28 & 62.23 & 2.60 & 52.30 & 1.46 & 56.25 & 1.81 \\ 
   58 & 57.55 & 1.74 & 59.76 & 1.97 & 59.98 & 2.31 & 62.92 & 2.63 & 52.97 & 1.49 & 56.99 & 1.84 \\ 
   59 & 58.30 & 1.77 & 60.56 & 2.02 & 60.61 & 2.38 & 63.59 & 2.68 & 53.62 & 1.54 & 57.71 & 1.88 \\ 
   60 & 59.04 & 1.81 & 61.35 & 2.06 & 61.22 & 2.41 & 64.24 & 2.71 & 54.27 & 1.57 & 58.42 & 1.92 \\ 
   61 & 59.76 & 1.85 & 62.12 & 2.09 & 61.83 & 2.45 & 64.88 & 2.73 & 54.91 & 1.60 & 59.13 & 1.94 \\ 
   62 & 60.48 & 1.90 & 62.88 & 2.12 & 62.42 & 2.50 & 65.50 & 2.78 & 55.55 & 1.65 & 59.82 & 1.98 \\ 
   63 & 61.19 & 1.93 & 63.63 & 2.17 & 63.01 & 2.52 & 66.12 & 2.78 & 56.18 & 1.67 & 60.51 & 2.02 \\ 
   64 & 61.89 & 1.96 & 64.37 & 2.19 & 63.60 & 2.57 & 66.73 & 2.81 & 56.82 & 1.71 & 61.19 & 2.03 \\ 
   65 & 62.60 & 2.01 & 65.10 & 2.21 & 64.19 & 2.59 & 67.34 & 2.82 & 57.45 & 1.75 & 61.88 & 2.06 \\ 
   66 & 63.30 & 2.03 & 65.84 & 2.23 & 64.79 & 2.62 & 67.95 & 2.82 & 58.10 & 1.77 & 62.58 & 2.09 \\ 
   67 & 64.02 & 2.05 & 66.57 & 2.26 & 65.39 & 2.65 & 68.56 & 2.84 & 58.76 & 1.80 & 63.27 & 2.11 \\ 
   68 & 64.74 & 2.09 & 67.31 & 2.27 & 66.01 & 2.66 & 69.18 & 2.83 & 59.43 & 1.84 & 63.99 & 2.12 \\ 
   69 & 65.47 & 2.12 & 68.06 & 2.26 & 66.64 & 2.69 & 69.80 & 2.82 & 60.13 & 1.86 & 64.71 & 2.14 \\ 
   70 & 66.21 & 2.13 & 68.81 & 2.27 & 67.28 & 2.70 & 70.44 & 2.82 & 60.85 & 1.89 & 65.46 & 2.16 \\ 
   71 & 66.98 & 2.14 & 69.58 & 2.28 & 67.95 & 2.70 & 71.09 & 2.79 & 61.59 & 1.93 & 66.23 & 2.16 \\ 
   72 & 67.77 & 2.16 & 70.36 & 2.26 & 68.65 & 2.72 & 71.77 & 2.77 & 62.38 & 1.96 & 67.03 & 2.16 \\ 
   73 & 68.59 & 2.17 & 71.17 & 2.23 & 69.37 & 2.72 & 72.46 & 2.75 & 63.21 & 1.97 & 67.86 & 2.16 \\ 
   74 & 69.43 & 2.17 & 71.99 & 2.20 & 70.13 & 2.70 & 73.17 & 2.69 & 64.09 & 1.99 & 68.73 & 2.16 \\ 
   75 & 70.32 & 2.16 & 72.84 & 2.17 & 70.93 & 2.68 & 73.92 & 2.63 & 65.02 & 2.01 & 69.64 & 2.14 \\ 
   76 & 71.24 & 2.13 & 73.71 & 2.13 & 71.78 & 2.66 & 74.69 & 2.58 & 66.03 & 2.03 & 70.60 & 2.12 \\ 
   77 & 72.20 & 2.10 & 74.61 & 2.08 & 72.66 & 2.62 & 75.49 & 2.52 & 67.11 & 2.04 & 71.61 & 2.08 \\ 
   78 & 73.21 & 2.06 & 75.55 & 2.02 & 73.60 & 2.57 & 76.33 & 2.43 & 68.27 & 2.04 & 72.67 & 2.03 \\ 
   79 & 74.27 & 2.01 & 76.51 & 1.94 & 74.58 & 2.49 & 77.20 & 2.33 & 69.52 & 2.03 & 73.79 & 1.96 \\ 
   80 & 75.37 & 1.94 & 77.50 & 1.85 & 75.62 & 2.40 & 78.10 & 2.22 & 70.86 & 1.99 & 74.96 & 1.88 \\ 
   81 & 76.51 & 1.86 & 78.53 & 1.76 & 76.70 & 2.29 & 79.03 & 2.10 & 72.29 & 1.94 & 76.18 & 1.79 \\ 
   82 & 77.70 & 1.75 & 79.58 & 1.65 & 77.83 & 2.16 & 80.00 & 1.97 & 73.82 & 1.86 & 77.45 & 1.69 \\ 
   83 & 78.92 & 1.63 & 80.65 & 1.53 & 79.00 & 2.02 & 81.00 & 1.84 & 75.42 & 1.76 & 78.76 & 1.57 \\ 
   84 & 80.17 & 1.51 & 81.75 & 1.41 & 80.20 & 1.87 & 82.02 & 1.70 & 77.08 & 1.63 & 80.10 & 1.43 \\ 
   85 & 81.44 & 1.39 & 82.86 & 1.29 & 81.42 & 1.72 & 83.06 & 1.57 & 78.77 & 1.49 & 81.46 & 1.31 \\ 
   86 & 82.72 & 1.26 & 83.99 & 1.18 & 82.66 & 1.56 & 84.12 & 1.43 & 80.47 & 1.34 & 82.82 & 1.17 \\ 
   87 & 84.01 & 1.13 & 85.13 & 1.06 & 83.92 & 1.39 & 85.20 & 1.29 & 82.16 & 1.18 & 84.19 & 1.04 \\ 
   88 & 85.29 & 1.01 & 86.27 & 0.96 & 85.18 & 1.24 & 86.29 & 1.16 & 83.82 & 1.03 & 85.54 & 0.93 \\ 
   89 & 86.58 & 0.90 & 87.43 & 0.86 & 86.44 & 1.10 & 87.40 & 1.03 & 85.43 & 0.90 & 86.88 & 0.81 \\ 
   90 & 87.87 & 0.79 & 88.59 & 0.76 & 87.70 & 0.96 & 88.52 & 0.91 & 87.01 & 0.77 & 88.21 & 0.71 \\ 
   91 & 89.15 & 0.69 & 89.75 & 0.66 & 88.98 & 0.82 & 89.66 & 0.79 & 88.54 & 0.67 & 89.52 & 0.62 \\ 
   92 & 90.45 & 0.59 & 90.94 & 0.57 & 90.27 & 0.70 & 90.82 & 0.67 & 90.03 & 0.56 & 90.82 & 0.52 \\ 
   93 & 91.77 & 0.49 & 92.14 & 0.46 & 91.60 & 0.57 & 92.02 & 0.54 & 91.52 & 0.46 & 92.11 & 0.42 \\ 
   94 & 93.15 & 0.35 & 93.40 & 0.34 & 93.01 & 0.41 & 93.29 & 0.39 & 93.02 & 0.33 & 93.42 & 0.30 \\ 
   95 & 94.61 & 0.17 & 94.73 & 0.16 & 94.54 & 0.19 & 94.67 & 0.18 & 94.57 & 0.16 & 94.74 & 0.14 \\ 
\bottomrule
\end{tabular}
\end{tiny}
\vspace{-0.5em}
\caption{\tiny Equated scores and \gls{see} for fall sessions using \gls{gke} and Sequential \gls{gke} across years.}
\label{tab:falls_multirow_years}
\end{table}
\end{document}